\documentclass[twocolumn,twocolappendix]{aastex631}

\usepackage{amsmath}
\usepackage{subfigure}
\providecommand{\sorthelp}[1]{}

\definecolor{dgreen}{RGB}{26,148,49}

\shorttitle{Mapping Galactic Dust Emission and Extinction with H\,{\sc I}, H\,{\sc II}, and H$_2$}
\shortauthors{Cheng et al.}
\graphicspath{{./}{figures/}}

\begin{document}

\title{Mapping Galactic Dust Emission and Extinction with H\,{\sc I}, H\,{\sc II}, and H$_2$}

\author[0000-0002-5437-0504]{Yun-Ting Cheng}
\affiliation{California Institute of Technology, 1200 E. California Boulevard, Pasadena, CA 91125, USA}
\affiliation{Jet Propulsion Laboratory, California Institute of Technology, 4800 Oak Grove Drive, Pasadena, CA 91109, USA}

\author[0000-0001-7449-4638]{Brandon S. Hensley}
\affiliation{Jet Propulsion Laboratory, California Institute of Technology, 4800 Oak Grove Drive, Pasadena, CA 91109, USA}

\author[0000-0001-5929-4187]{Tzu-Ching Chang}
\affiliation{Jet Propulsion Laboratory, California Institute of Technology, 4800 Oak Grove Drive, Pasadena, CA 91109, USA}
\affiliation{California Institute of Technology, 1200 E. California Boulevard, Pasadena, CA 91125, USA}

\author[0000-0001-7432-2932]{Olivier Dor{\'e}}
\affiliation{Jet Propulsion Laboratory, California Institute of Technology, 4800 Oak Grove Drive, Pasadena, CA 91109, USA}
\affiliation{California Institute of Technology, 1200 E. California Boulevard, Pasadena, CA 91125, USA}

\begin{abstract}
Neutral hydrogen (\ion{H}{1}) emission closely traces the dust column density at high Galactic latitudes and is thus a powerful tool for predicting dust extinction. However, the relation between \ion{H}{1} column density $N_{\rm HI}$ and high-latitude dust emission observed by Planck has large-scale residuals at the level of $\lesssim 20\%$ on tens of degree scales. In this work, we improve \ion{H}{1}-based dust templates in the north/south Galactic poles covering a sky fraction of $f_{\rm sky}=13.5\%/11.0\%$ (5577/4555\,deg$^2$) by incorporating data from ionized (\ion{H}{2}) and molecular (H$_2$) gas phases. We make further improvements by employing a clustering analysis on the \ion{H}{1} spectral data to identify discrete clouds with distinct dust properties. However, only a modest reduction in fitting residuals is achieved. We quantify the contributions to these residuals from variations in the dust-to-gas ratio, dust temperature and opacity, and magnetic field orientation using ancillary datasets. Although residuals in a few particular regions can be attributed to these factors, no single explanation accounts for the majority. Assuming a constant dust temperature along each line of sight, we derive an upper limit on the high-latitude dust temperature variation of $\sigma_T<1.28$K, lower than the temperature variation reported in previous studies. Joint analysis of multiple existing and upcoming datasets that trace Galactic gas and dust properties is needed to clarify the origins of the variation of gas and dust properties found here and to significantly improve gas-based extinction maps.\footnote{\textcircled{c} 2025. All rights reserved.
}
\end{abstract}


\section{Introduction} \label{S:intro}

Neutral hydrogen (\ion{H}{1}) emission traces many properties of dust in the interstellar medium (ISM). The strong coupling between gas and dust renders the \ion{H}{1} column density an excellent proxy for the dust column density---and thus for both dust emission and extinction \citep{1996A&A...312..256B, planck2013-p06b, 2017ApJ...846...38L}---on diffuse lines of sight. The filamentary structure of \ion{H}{1} is elongated along the direction of the local magnetic field, enabling \ion{H}{1} emission data alone to predict properties of polarized dust emission, including polarized intensity, polarization angle, and degree of polarization \citep{Clark:2015, Clark:2018, 2019ApJ...887..159H}. The velocity information afforded by spectroscopic \ion{H}{1} maps encodes line-of-sight information, allowing many of these properties to be mapped in a three-dimensional way \citep{2019ApJ...887..136C}.

\ion{H}{1} emission is particularly well suited for tracing dust in a cosmological context. The strongest coupling between \ion{H}{1} and dust emission is at low column densities ($N_{\rm HI} \lesssim 4\times10^{20}\,$cm$^{-2}$) where \ion{H}{1} is optically thin and there is little molecular gas, precisely the regions targeted by most cosmological surveys. Further, since they are based on a bright spectroscopic line, \ion{H}{1} maps are essentially free of extragalactic contamination that could bias cosmological inferences \citep{2019ApJ...870..120C}. Thus, \ion{H}{1} maps have seen use in cosmological contexts as a tracer of dust reddening (i.e., differential extinction), emission, and polarization \citep[e.g.,][]{planck2013-p03f, planck2013-p13, 2019ApJ...883...75L, Ade:2023}.

Currently, the most widely used dust reddening map is that of \citet[][hereafter SFD]{1998ApJ...500..525S}. The SFD map was constructed from the 100\,$\mu$m IRAS map \citep{1984ApJ...278L...1N}, which is dominated by the thermal emission from dust and thus closely correlated with dust extinction. A dust temperature correction was made using the 100 and 240\,$\mu$m maps from the Diffuse Infrared Background Experiment (DIRBE) instrument on the Cosmic Background Explorer \citep[COBE; ][]{1992ApJ...397..420B} since the same column density of dust produces more emission at shorter wavelengths at higher temperatures. However, the SFD map is imperfect. It is known to contain imprints from the cosmic infrared background \citep[CIB;][]{2007PASJ...59..205Y,2019ApJ...870..120C}. Large-scale discrepancies have been observed relative to direct optical reddening measurements \citep{2010ApJ...719..415P} and \ion{H}{1} maps \citep{2017ApJ...846...38L} that are unlikely to be caused by the CIB.

The demands of ongoing and upcoming cosmological surveys motivate the need to improve the fidelity of available reddening maps and quantify uncertainties based on different methods and reddening tracers \citep{2012MNRAS.424..564R,2013MNRAS.432.2945H,2020MNRAS.496.2262K,2022JCAP...07..041C,2023AJ....165...58Z,2024arXiv240815909K}. For example, a cross-correlation analysis of DESI emission line galaxy samples and the CMB lensing found that the $\sigma_8$ tension with the Planck value changes from 4.29$\sigma$ to 3.07$\sigma$ when using different reddening templates for systematic correction \citep{2024arXiv240815909K}. One option for improving reddening maps is to remove the extragalactic signal from the SFD map, as done recently through cross-correlation with extragalactic galaxy catalogs \citep{2023arXiv230603926C}. However, CIB contamination is limited to relatively small angular scales, and so this approach does not address sources of systematic errors that affect the SFD map on scales of many degrees. 

In this work, we use gas-based tracers of the dust column density to construct extinction maps independent of dust emission and to quantify these uncertainties. Previous analyses \citep[e.g.,][]{2017ApJ...846...38L} have generally employed only \ion{H}{1} as the gas tracer and used simple velocity cuts on \ion{H}{1} spectral data to exclude contributions from dust-poor high-velocity clouds (HVCs). This does not fully leverage information from the \ion{H}{1} distribution in both position and velocity space. With the availability of \ion{H}{1} maps with high spectral resolution and sensitivity, along with recent developments in mapping ionized (\ion{H}{2}) and molecular (H$_2$) gas phases, we improve existing \ion{H}{1}-based extinction maps in two aspects.

First, we employ a multi-phase analysis that incorporates information from \ion{H}{2} and H$_2$ gas to trace dust associated with gas not in the neutral atomic phase. Second, we utilize an unsupervised clustering algorithm to identify distinct large-scale \ion{H}{1} structures from spectral maps, which may be associated with different dust properties. We produce a new dust extinction template by incorporating both of the aforementioned improvements: multi-phase gas tracers and \ion{H}{1} components derived from a clustering-based method.

Ultimately, the resulting dust template leads to only minor improvements in the residuals when fitting our templates to far-infrared (FIR) dust emission maps from Planck relative to fitting a simple velocity-filtered \ion{H}{1} map. We therefore use ancillary datasets to quantify the contributions of dust-to-gas ratio variations, dust temperature variations, and magnetic field orientation effects to the remaining mismatch between \ion{H}{1} and dust emission. No single effect dominates, suggesting that incorporation of additional datasets, such as stellar extinction maps and dust polarization maps, will be required to further improve these gas-based models. We make our high Galactic latitude extinction map, covering a sky fraction of $f_{\rm sky}=13.5\%/11.0\%$ (5577/4555\,deg$^2$) in the North/South Galactic Cap (NGC/SGC), publicly available.

This paper is organized as follows: in Section~\ref{S:data}, we describe the datasets used in this analysis. Section~\ref{S:model_fitting} presents the theoretical framework of our dust model built from the gas templates. Sections~\ref{S:multiphase} and~\ref{S:clustering_based_model} detail our improved modeling approach, which includes multi-phase analysis and a clustering-based algorithm for defining distinct \ion{H}{1} structures, respectively. Our final model is presented in Section~\ref{S:combined_model}. In Section~\ref{S:interpretation}, we explore different possibilities for the origins of the residuals that remain in our model fit. We produce a new dust reddening map using our updated templates, with details described in Section~\ref{S:data_release_sfd}. Finally, we conclude in Section~\ref{S:conclusions}.

\section{Data}\label{S:data}
The goal of this work is to relate maps of multi-phase gas emission (\ion{H}{1}, \ion{H}{2}, and H$_2$) to maps of FIR emission from Galactic dust. This section details the datasets used in our analysis. 

The data used in this work have a range of angular resolutions and are provided at various HEALPix\footnote{\url{http://healpix.sf.net}} \citep{2005ApJ...622..759G} pixelizations. As we focus on large-scale dust structures, we perform all analyses with $N_{\rm side}=128$ pixelization, corresponding to $\sim 0.5^\circ$ resolution, much coarser than the native resolution of all of the datasets.

\subsection{\texorpdfstring{\ion{H}{1} Data}{HI Data}}
We use the all-sky \ion{H}{1} spectral survey data from the HI4PI survey \citep{2016A&A...594A.116H}, which combines data from the Effelsberg-Bonn \ion{H}{1} Survey \citep[EBHIS; ][]{2010ApJS..188..488W,2011AN....332..637K,2016A&A...585A..41W} and the Galactic All-Sky Survey \citep[GASS; ][]{2009ApJS..181..398M,2010A&A...521A..17K,2015A&A...578A..78K}. The HI4PI dataset has an FWHM angular resolution of $16.'2$ and a spectral resolution of 1.49\,km\,s$^{-1}$, covering the velocity range $|v|<600$\,km\,s$^{-1}$. 

Since our analysis focuses on the large-scale morphology of the \ion{H}{1} and does not require fine binning in \ion{H}{1} velocity, we use the HI4PI data products\footnote{\url{https://dataverse.harvard.edu/dataset.xhtml?persistentId=doi:10.7910/DVN/P41KDE}} of \citet{2019ApJ...887..136C}, who binned the raw HI4PI data into 41 velocity channels. In their binning scheme, the bin widths increase toward higher velocities within $|v|\lesssim 90$\,km\,s$^{-1}$. We convert the HI4PI map from units of K\,km\,s$^{-1}$ to \ion{H}{1} column density in cm$^{-2}$ using the conversion factor $1.82\times 10^{18}$, under the assumption that the \ion{H}{1} is in the optically thin regime \citep{2019ApJ...887..136C}. The \ion{H}{1} spectral maps are provided in the HEALPix format with $N_{\rm side}=1024$.

\subsection{\texorpdfstring{\ion{H}{2} Data}{HII Data}}\label{S:H2_data}
Most of the free electrons in the diffuse ISM are from ionized hydrogen atoms, and so the electron column density can serve as a proxy for the column density of \ion{H}{2}. The ``dispersion measure'' (DM), a direct measure of the electron column density, can be inferred along sight lines with pulsars, where the deterministic frequency dependence of the pulsation signal arrival time is related to the DM. However, pulsar DM measurements are sparse, and thus additional observables are required to build a complete DM map of the sky. Other indirect tracers of the electron column include the ``rotation measure'' (RM), which determines the spectral dependence of the Faraday rotation of photons. RM is proportional to the electron column density weighted by the line-of-sight magnetic field. Additionally, one can constrain the column of the square of the electron density along the line of sight, known as the ``emission measure'' (EM), which is proportional to the intensity of free-free emission and H$\alpha$ emission.

Combining multiple observables of DMs, RMs and EMs, \citet{2024A&A...690A.314H} uses a Bayesian inference framework to construct all-sky maps of DM, EM, and the Galactic line-of-sight magnetic field. We adopt their DM map\footnote{\url{https://zenodo.org/records/10523170}} directly as a map of the \ion{H}{2} column density. The DM map and its variance map are provided at $N_{\rm side}=256$.

\subsection{\texorpdfstring{$H_2$ Data}{H2 Data}}
Some of the gas even at high Galactic latitudes is in molecular form \citep[H$_2$,][]{Magnani2017}. However, emission from H$_2$ itself is not readily observable. Instead, the column density of H$_2$ ($N_{\rm H{_2}}$) can be estimated from CO emission by assuming a constant scaling factor ($X_{\rm CO}$) between the CO(1$-$0) brightness temperature ($W_{\rm CO}$) and the column density of hydrogen atoms in the form of H$_2$, i.e., $N_{\rm H{_2}}=2X_{\rm CO}W_{\rm CO}$. Here, we adopt the conversion factor $X_{\rm CO}=2\times10^{20}$~cm$^{-2}$~K$^{-1}$~km$^{-1}$~s \citep{2013ARA&A..51..207B}. However, we note that our H$_2$ template does not trace the CO-dark H$_2$, which may constitute a non-negligible gas column in high-latitude regions. See Section~\ref{S:interp_gtdr} for further discussion.

Full-sky spectroscopic surveys of CO emission are challenging to undertake. However, CO can contribute non-negligible emission even to broad photometric bands, and so full-sky maps of CO emission have been derived from the multi-frequency Planck data \citep{planck2013-p03a, planck2014-a12}. Recently, \citet{2024A&A...688A..54G} presented a reprocessing of the Planck CO maps with an extension of the Generalized Needlet Internal Linear Combination method to derive a set of full-sky CO maps with reduced noise. We adopt their CO(1-0) map and its corresponding uncertainty maps\footnote{\url{https://portal.nersc.gov/project/cmb/Planck_Revisited/co/}}. The maps are released at $N_{\rm side}=1024$ with an angular resolution of $\sim 10'$ (their apodization scale). Systematic and statistical errors are quantified separately. We obtain the systematic uncertainties directly from the provided systematic error map and obtain the statistical uncertainties from the variance of the provided 100 noise simulations. We then calculate the total uncertainties in the CO(1$-$0) map by summing the systematic and statistical error maps in quadrature.

\subsection{Planck Maps}\label{S:Planck_maps}
We focus our analysis on the three highest-frequency Planck bands (353, 545, and 857\,GHz), which contain less emission from the CMB, synchrotron, and free-free emission than the lower-frequency maps. We employ maps from the Planck Public Release~3 \citep[PR3;][]{planck2016-l03} that have been processed by subtracting CMB fluctuations \citep{2019ApJ...883...75L}. The resulting maps\footnote{\url{https://dataverse.harvard.edu/dataset.xhtml?persistentId=doi:10.7910/DVN/8A1SR3}} are provided at $N_{\rm side} = 2048$ with a resolution of $\sim$5$'$. We use the covariance matrix from the original Planck PR3 dataset\footnote{Specifically, the \texttt{II\_COV} field in, e.g., \texttt{HFI\_SkyMap\_545\_2048\_R3.01\_full.fits} as the pixel variance in intensity.}.

\subsection{Masks}\label{S:masks}
It has been demonstrated that for $N_{\rm H\text{\sc I}} < 4\times10^{20}$\,cm$^{-2}$ there exists a strong linear correlation between \ion{H}{1} and dust extinction \citep{planck2013-p06b, 2017ApJ...846...38L}. Thus, we focus on diffuse, high Galactic latitudes in this work.

We adopt the masks\footnote{Denoted \texttt{4.0e 20\_gp40.7z}} from \citet{2019ApJ...883...75L} that incorporate a threshold $N_{\rm H\text{\sc I}}=4 \times 10^{20}$\,cm$^{-2}$, a $40\%$ Galactic plane mask from Planck, and pixels that contain extragalactic point sources from \citet{planck2014-a35}. However, this mask also removes sight lines with molecular intermediate-velocity clouds and CO-dark molecular gas from a census of these objects \citep{2016A&A...596A..94R}, as well as regions with residual emission identified in their process of fitting Planck maps to \ion{H}{1}. The Galactic plane and the $N_{\rm H\text{\sc I}}$ threshold mask account for most of the masked pixels and the large-scale geometry of the mask, and the remaining masking criteria only remove some additional small, compact regions. While our analysis aims to explore the dust-gas relationship across different phases, including molecular clouds, this additional masking covers only a small fraction of the sky and does not affect our analysis of large-scale features over much of the diffuse high-latitude sky. A separate mask is provided for each of the three frequency bands (353, 545, and 857\,GHz), and we take the union of them as our final common mask for all three Planck bands. We use the Boolean (non-apodized) version of the mask.

After applying this mask, we are left with 5577/4555\,deg$^2$ ($13.5\%/11.0\%$ of the sky) in the NGC/SGC, respectively.

\section{Model Fitting}\label{S:model_fitting}

\subsection{Theoretical Framework}\label{S:theory}
Dust and gas are well coupled in the ISM: the total H column density $N_{\rm H}$ (i.e., summing over neutral atomic, ionized atomic, and molecular H) is an excellent proxy for the dust mass surface density $\Sigma_d$ \cite[e.g.,][]{Bohlin:1978}. We define the dust to gas mass ratio $\delta_{\rm DG}$ as

\begin{equation}
    \delta_{\rm DG} \equiv \frac{\Sigma_d}{m_p N_{\rm H}}
    ~~~,
\end{equation}
where $m_p$ is the proton mass. The Milky Way has $\delta_{\rm DG} \sim 0.01$ \citep{Draine:2021}.

At FIR wavelengths, the observed dust emission is dominated by grains with steady-state temperatures \citep{Li:2001}. A population of dust with mass surface density $\Sigma_d$ at temperature $T$ and opacity $\kappa_\nu$ at frequency $\nu$ emits radiation with specific intensity
\begin{equation} \label{eq:theory}
\begin{split}
    I_\nu &= \Sigma_d \kappa_\nu B_\nu\left(T\right) \\
    &= N_{\rm H} m_p \delta_{\rm DG} \kappa_\nu B_\nu\left(T\right)
    ~~~,
\end{split}
\end{equation}
where $B_\nu\left(T\right)$ is the Planck function
\begin{equation}\label{E:MBB}
B_\nu\left(T\right) = \frac{2h\nu^3}{c^2}\frac{1}{e^{h\nu/k_B T}-1} ~~~,
\end{equation}
$c$ is the speed of light, $h$ is the Planck constant, and $k_B$ is the Boltzmann constant. Here we have neglected the dependence of the observed intensity on the orientation of the local magnetic field, arising from the fact that the opacity of a population of aligned aspherical grains is orientation-dependent \citep{Lee:1985}. We return to this effect in Section~\ref{S:polarization}.

The dust emissivity per $N_{\rm H}$ is defined as
\begin{equation}\label{E:emissivity}
    \epsilon_\nu \equiv m_p \delta_{\rm DG} \kappa_\nu B_\nu\left(T\right)
\end{equation}
such that $I_\nu = \epsilon_\nu N_{\rm H}$. The dust emissivity is commonly described by a modified blackbody (MBB) model:
\begin{equation}
\epsilon_\nu~~\propto~~\nu^{\beta}B_\nu(T)\\
\end{equation}
where the dust opacity $\kappa_\nu$ follows a power-law frequency dependence with the spectral index $\beta$. While our analysis fits for emissivity $\epsilon_\nu$ in each band independently, we provide more discussion on the MBB parameters $T$ and $\beta$ in Section~\ref{S:T_var}.

The distribution of dust and gas in the Galaxy is not uniform either along the line of sight or in the plane of the sky. Rather, dust and gas appear organized into clumps and filaments, structures we will refer to simply as ``clouds.'' The principal assumption of this work is that each cloud $i$ in the gas phase $x$ ($x\in\{$\ion{H}{1}, \ion{H}{2}, H$_2$$\}$) can be characterized by an emissivity $\epsilon_{\nu, i}^x$ that is constant across the extent of the cloud. This modeling has proven effective in predicting dust polarization properties from the 3D position-position-velocity (PPV) of \ion{H}{1} \citep[e.g.,][]{2019ApJ...887..136C, Pelgrims:2021, 2024ApJ...972...66L}. 

In this work, we separate \ion{H}{1} into multiple clouds with their structures in the PPV space. For \ion{H}{2} and H$_2$, only 2D column density maps are available. Given that the ionized and molecular phases are subdominant in the regions of sky we consider in this work, and the fact that we do not have velocity information for these phases, we employ a single emissivity for all \ion{H}{2} and H$_2$ gas, respectively. Hereafter, $\epsilon_{\nu,i}$ refers to $\epsilon_{\nu,i}^{{\rm HI}}$, the emissivity of the $i$th \ion{H}{1} cloud, and $\epsilon_{\nu,{\rm HII}}$ and $\epsilon_{\nu,{\rm H}_{2}}$ denote the emissivity of \ion{H}{2} and H$_2$ phases, respectively.

Summing over discrete clouds, the total observed dust emission on a single line of sight at frequency $\nu$ is

\begin{equation}\label{E:linear_template_model}
I_\nu(\hat{\theta}) = \left[\sum_{i}^{N_c} \epsilon_{\nu, i} N_{{\rm HI}, i}(\hat{\theta})\right] + \epsilon_{\nu,{\rm HII}} N_{{\rm HII}}(\hat{\theta}) + 2\epsilon_{\nu,{\rm H}_{2}} N_{{\rm H}_{2}}(\hat{\theta})
\end{equation}
where $\hat{\theta}$ is the angular position on the sky, $N_c$ is the number of clouds along the line of sight, $N_{{\rm HI}, i}$ is the \ion{H}{1} column density associated with cloud $i$, and $N_{{\rm HII}}$ and $N_{{\rm H}_{2}}$ are the \ion{H}{2} and H$_2$ column densities, respectively. The factor of 2 in the H$_2$ term accounts for the fact that one molecule of H$_2$ contains two H atoms.

With this formalism in place, the underlying questions motivating this work are:
\begin{enumerate}
    \item Does the dust emissivity differ across gas phases? Can previous high-latitude dust modeling that relies only on \ion{H}{1} information be improved with multi-phase information?
    \item Are there large-scale, coherent \ion{H}{1} structures on the sky identifiable in PPV space each with their own $\epsilon_\nu$ that explain the observed large-scale residuals when fitting \ion{H}{1} data to dust emission data?
    \item Are there large-scale, coherent variations in $\delta_{\rm DG}$, $T_d$, and/or $\kappa_\nu$?
    \item Can ancillary data be employed to constrain variations in these parameters to identify how dust and gas properties are varying in the Galactic ISM?
\end{enumerate}

\subsection{Analysis Framework}
We model the Planck intensity map at frequency $\nu$ with a linear model of multi-phase hydrogen column templates, $\{N_{{\rm H\text{\sc I}},i},N_{{\rm H\text{\sc II}}},N_{{\rm H}_2}\}$, as outlined in Equation~\eqref{E:linear_template_model}. This yields the parametric model:
\begin{equation}\label{E:HI_lin_fit}
\begin{split}
I^m_\nu(\nu,\hat{\theta};\{\epsilon_{\nu}\},b_\nu) =& \left[\sum_i^{N_c} \epsilon_{\nu,i} N_{{\rm H}\text{\sc I},i}(\hat{\theta})\right] \\
&+ \epsilon_{\nu,{\rm HII}} N_{{\rm HII}}(\hat{\theta}) \\
&+ 2\epsilon_{\nu,{\rm H}_{2}} N_{{\rm H}_{2}}(\hat{\theta}) + b_\nu~~~,
\end{split}
\end{equation}
where the offset term $b_\nu$ is a nuisance parameter accounting for zero-point values in the Planck maps and the template maps.

With a set of templates of \ion{H}{1} clouds $N_{{\rm HI},i}$ and the column density map of $N_{{\rm HII}}$ and $N_{{\rm H}_2}$, we fit our model $I^m_\nu(\nu,\hat{\theta}_j;\{\epsilon_{\nu}\},b_\nu)$ to the Planck data $I^d_\nu(\nu,\hat{\theta}_j)$, where $j$ denotes the pixel index. The best-fit values of the emissivities $\left \{ \epsilon_{\nu}\right \}$ and the offset $b_\nu$ are determined by minimizing the $\chi^2$ value
\begin{equation}\label{E:chi2}
\chi^2=\sum_{j=1}^{N_{\rm pix}}\frac{\left(I^d_\nu(\nu,\hat{\theta}_j)-I^m_\nu(\nu,\hat{\theta}_j;\{\epsilon_{\nu}\},b_\nu)\right)^2}{\sigma^2(\nu, \hat{\theta}_j)}~~~,
\end{equation}
where $\sigma^2(\nu, \hat{\theta}_j)$ is the noise variance of the $j$th pixel in the Planck maps at frequency band $\nu$ and $N_{\rm pix}$ is the total number of pixels in the map.

We fit each of the three Planck bands separately for both the NGC and the SGC regions. The coefficients of our linear model ($\epsilon_{\nu}$'s and $b_\nu$) for a given H template set can be obtained using the closed-form solution
\begin{equation}\label{E:ci_matrix}
\hat{\epsilon}_\nu=(\mathbf{A}^T\mathbf{N}^{-1}\mathbf{A})^{-1}\mathbf{A}^T\mathbf{N}^{-1}d~~~,
\end{equation}
where $\hat{\epsilon}_\nu=(\epsilon_{\nu,1}, \epsilon_{\nu,2},...,\epsilon_{\nu,{N_c}}, \epsilon_{\nu,{\rm HII}}, \epsilon_{\nu,{\rm H}_{2}}, b_\nu)$, $d$ is the 1D flattened $N_{\rm pix}$-sized data map vector, $\mathbf{A}$ is an $N_{\rm pix}\times (N_c+3)$ matrix where its first $N_c+2$ columns are the flattened $N_{{\rm HI},i}$, $N_{{\rm HII}}$ and $N_{{\rm H}_{2}}$ maps, and the elements of the last column are all ones for the offset $b_\nu$. $\mathbf{N}$ is the noise covariance matrix, taken to be an $N_{\rm pix}\times N_{\rm pix}$ diagonal matrix with the flattened noise variance $\sigma^2$ as its diagonal element. In reality, there is pixel covariance arising from both correlated instrument noise and the clustering of the CIB. We discuss this further in Section~\ref{S:combined_model}, where we find consistency with other analyses that employed the full covariance matrix.

\subsection{Previous Models}\label{S:previous_model}
\ion{H}{1} has been used to build Galactic dust emission or extinction templates in several previous analyses. For example, a velocity cut on \ion{H}{1} data was used to build Galactic dust emission templates for analysis of the CIB with Planck data \citep{planck2013-p13}. They used a two-component \ion{H}{1} template set consisting of a low-velocity cloud (LVC) component, defined by $|v|<30$\,km\,s$^{-1}$ and an intermediate velocity cloud (IVC) component, defined by $30<|v|<90$\,km\,s$^{-1}$. They then fit the (spatially varying) emissivity of LVC and IVC templates to Planck data in pixels of size 13\,deg$^{2}$ (i.e., $N_{\rm side}=16$). \citet{2017ApJ...846...38L} investigated the correlation between the \ion{H}{1} column density and dust extinction and built a dust extinction template at high Galactic latitude using a single \ion{H}{1} column density map defined by $|v|<90$\,km\,s$^{-1}$.

\begin{figure*}[ht!]
\begin{center}
\includegraphics[width=0.48\linewidth]{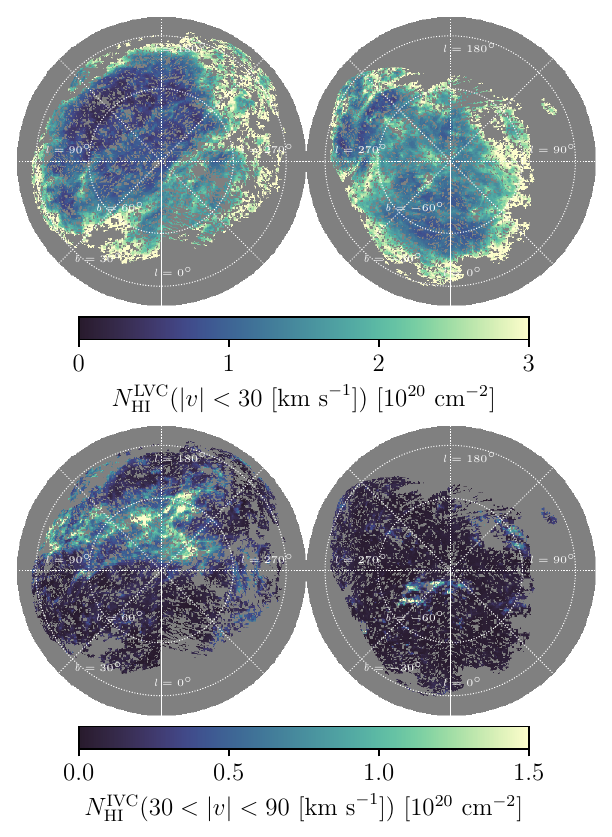}
\includegraphics[width=0.48\linewidth]{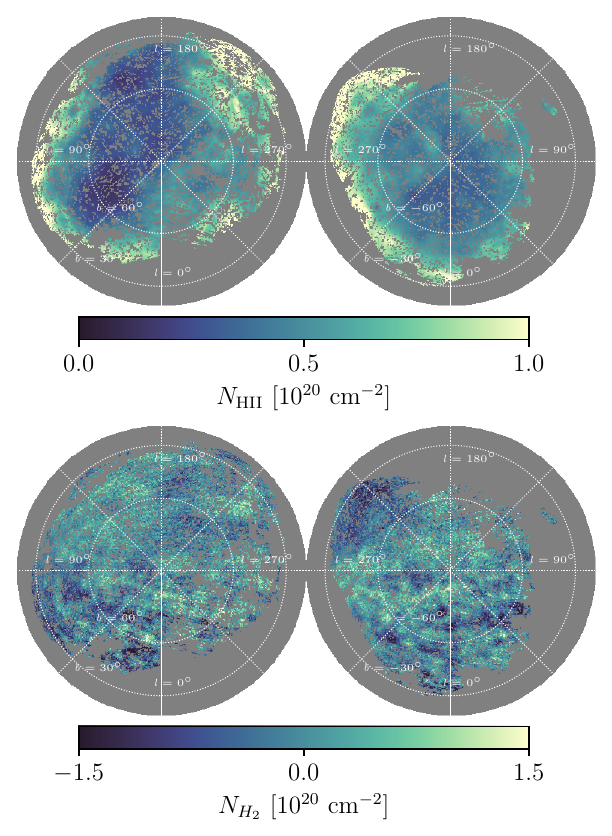}
\caption{\label{F:H_phases_maps} Column density maps of various gas components, presented in orthographic projection centered on the north/south Galactic poles. The left/right panels display the NGC/SGC, respectively, with gray regions indicating the masked areas as described in Section~\ref{S:masks}. Top-left panel: \ion{H}{1} column density in the LVC component defined by $|v|<30$\,km\,s$^{-1}$. Bottom-left panel: \ion{H}{1} column density in the IVC component defined by $30<|v|<90$\,km\,s$^{-1}$. Top-right panel: H column density in the ionized phase (\ion{H}{2}). Bottom-right panel: H column density in the molecular phase (H$_2$), derived from the CO map with a constant CO-to-H$_2$ ratio $X_{\rm CO}$: $N_{\rm H_2}=2X_{\rm CO}W_{\rm CO}$. The CO map is noise-dominated, and thus the pixel value fluctuates around zero.}
\end{center}
\end{figure*}

Figure~\ref{F:H_phases_maps} shows the LVC and IVC maps with the velocity ranges defined in \citet{planck2013-p06b} from the HI4PI survey. The \ion{H}{1} template in \citet{2017ApJ...846...38L} is the sum of the LVC and IVC maps. We take these one- and two-component \ion{H}{1}-only models as the baseline to compare with our new model presented in Section~\ref{S:clustering_based_model}.

\begin{figure*}[ht!]
\begin{center}
\includegraphics[width=\linewidth]{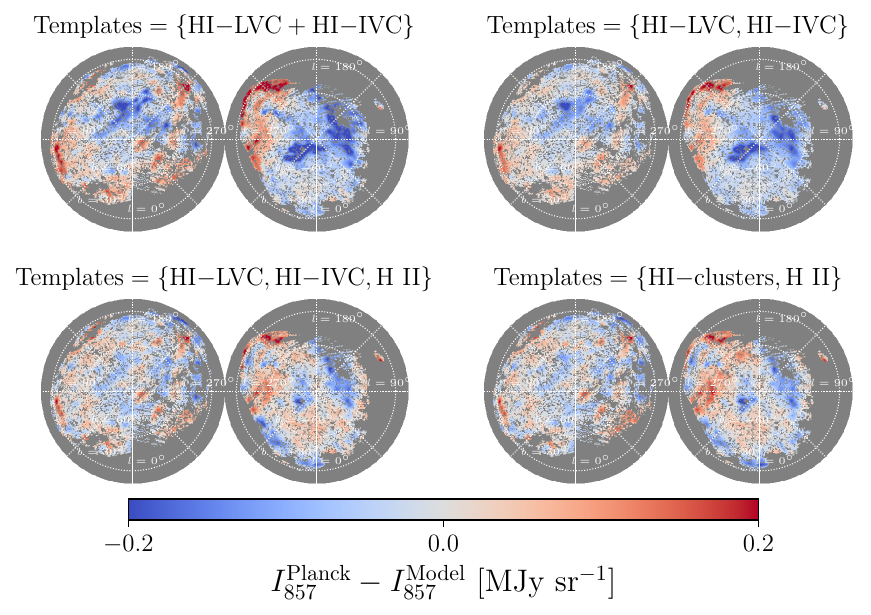}
\caption{\label{F:clus_res_maps} Residuals of a linear fit to the Planck 857\,GHz map with different combinations of H templates. Top left panel: the total \ion{H}{1}, i.e. the sum of the LVC (top-left panel in Figure~\ref{F:H_phases_maps}) and the IVC (bottom-left panel in Figure~\ref{F:H_phases_maps}) components; Top right panel: the LVC and the IVC of \ion{H}{1}; Bottom left panel: the LVC and the IVC of \ion{H}{1} and the \ion{H}{2} map (top-right panel in Figure~\ref{F:H_phases_maps}); Bottom right panel: our \ion{H}{1} templates constructed with clustering algorithm (Section~\ref{S:clustering_based_model}) and the \ion{H}{2} map. In all of the fitting, a constant offset level ($b_\nu$) is jointly fit with the templates. The NGC and the SGC are fitted separately. All of the residual maps are smoothed with a 1$^\circ$ Gaussian kernel to highlight the large-scale patterns.}
\end{center}
\end{figure*}

\begin{figure}[ht!]
\begin{center}
\includegraphics[width=\linewidth]{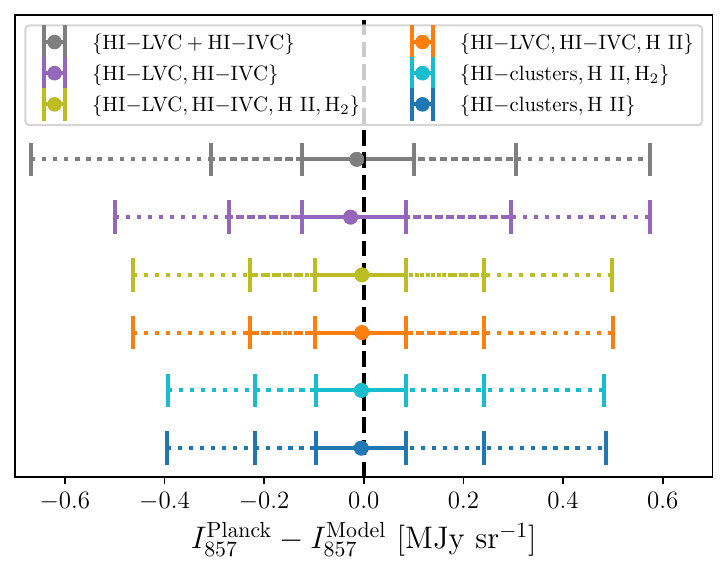}
\caption{\label{F:clus_res_sigmas} 1$\sigma$, 2$\sigma$, and 3$\sigma$ intervals of residuals of linear fits to the Planck 857\,GHz map with different combinations of H templates shown in Figure~\ref{F:clus_res_maps}, as well as two additional cases that incorporate the H$_2$ template into the fitting. We see that adding H$_2$ makes negligible changes in the residual distribution.}
\end{center}
\end{figure}

We fit a linear model to the Planck 857\,GHz band using Equation~\eqref{E:HI_lin_fit} with both the LVC and IVC of \ion{H}{1} as the two-component template set, as well as the total \ion{H}{1} (LVC$+$IVC) as a single-component template. The residual maps, defined as the Planck data map minus the model map, and the median, 1$\sigma$, 2$\sigma$, and 3$\sigma$ intervals of the distribution are shown in Figures~\ref{F:clus_res_maps} and \ref{F:clus_res_sigmas}. We observe that in the single-component fitting case, the residuals are negative in places with strong IVC emission. This error is mitigated in the two-component fitting, suggesting that IVCs emit less FIR radiation per H atom than LVCs, and thus have a lower emissivity $\epsilon_{\nu}$ than LVCs, consistent with previous findings \citep[e.g.,][]{planck2011-7.12, 2017ApJ...846...38L}.
Furthermore, without separately fitting LVCs and IVCs, the residuals exhibit a large negative tail beyond the $\sim 2\sigma$ limit, whereas these outliers are greatly mitigated by having separate LVC and IVC templates. Therefore, it is essential to separate the LVC and IVC components in the model.

While previous studies have found that \ion{H}{1} alone provides a good template for high-latitude dust emission in Planck, there are still residuals in the fit that exhibit large-scale patterns. Both the omission of H atoms in other phases and the simple velocity cut for separating low- and high-velocity \ion{H}{1} clouds may contribute to these residuals. Therefore, in this study, we aim to improve the previous \ion{H}{1}-based dust model by (1) employing a multi-phase analysis that incorporates \ion{H}{2} and H$_2$ templates along with the \ion{H}{1}, and (2) refining the algorithm for separating \ion{H}{1} emission into discrete structures using a data-driven method. To assess the relative improvements from incorporating each factor individually, we first examine the model improvement from (1) in Section~\ref{S:multiphase}, and then add (2) in Section~\ref{S:clustering_based_model}. Our final model, which includes both (1) and (2), is presented in Section~\ref{S:combined_model}.

\section{Multi-phase Analysis}\label{S:multiphase}
In this section, we examine the improvement to an \ion{H}{1}-only model gained by incorporating \ion{H}{2} and/or H$_2$ templates. Since dust can reside both in ionized and molecular gas, \ion{H}{1} alone likely underestimates the dust column density. With \ion{H}{2} and H$_2$ maps now available, we conduct a multi-phase fitting by incorporating them into our template set.

The \ion{H}{2} and H$_2$ maps are displayed in Figure~\ref{F:H_phases_maps}. We add the \ion{H}{2} and H$_2$ templates one at a time and also both together, in combination with the LVC and IVC \ion{H}{1} templates. For every combination of template sets, we calculate the emissivity associated with each template component $\epsilon_\nu$ and the offset $b_\nu$ with Equation~\eqref{E:ci_matrix}. 

We find the inclusion of the \ion{H}{2} template in the fit reduces the number of pixels with large residuals (see Figure~\ref{F:clus_res_sigmas}). Thus, the \ion{H}{2} template is predictive of dust emission not traced by \ion{H}{1}, although the relative improvement of the fit with versus without the \ion{H}{2} template is small. This finding is in agreement with previous studies that find dust emission associated with the ionized gas. For example, \citet{2000A&A...354..247L} found that dust emission associated with the warm ionized medium (WIM) at high Galactic latitude has a similar temperature and emissivity to that associated with the neutral phase, using data from DIRBE and FIRAS and ionized gas traced by H$\alpha$ emission. A recent study has also claimed detection of FIR dust emission associated with the WIM at high Galactic latitude sight lines using the DMs of pulsars \citep{2023ApJ...959..115W}. 

We caution, however, that the \ion{H}{2} template exhibits a high correlation with the total dust column. We find a Pearson correlation coefficient of $r\sim0.5$ between our \ion{H}{2} template and the dust reddening map from \citet{1998ApJ...500..525S}. This indicates that, in addition to tracing \ion{H}{2}, our template also captures any components that correlate with the total dust and gas column. These include H$_2$ gas, either in the CO-dark phase or with CO emission below the sensitivity limit of our CO template. \citet{2006ApJ...636..908G} found that the molecular hydrogen fraction at some high-latitude sight lines can reach up to $\sim30\%$ even at column densities as low as $2\times10^{20}$~cm$^{-2}$, suggesting that some H$_2$ may be absorbed into our \ion{H}{2} template fit. Furthermore, the strong correlation between the \ion{H}{2} and \ion{H}{1} templates introduces degeneracy in the fit. In our final model, the inferred dust emissivity in ionized gas ranges from roughly half of that in \ion{H}{1} to comparable values (see Table~\ref{T:params}), likely hampered by degeneracy with the \ion{H}{1} templates and the presence of molecular gas.

On the other hand, we find adding H$_2$ to the \ion{H}{1} templates makes a negligible difference in the residual map (see Figure~\ref{F:clus_res_sigmas}). When performing a fit with all of the \ion{H}{1}, \ion{H}{2}, and H$_2$ templates simultaneously, we find that the fitted value of H$_2$ emissivity is negligibly small compared to other the components. Including or excluding the H$_2$ template has almost no impact on the fit parameters (emissivities and offsets) of the other components. While the residuals of fitting Planck data with the template set $\{$\ion{H}{1}-LVC, \ion{H}{1}-IVC$\}$ have a $\sim 25\%$ correlation coefficient with \ion{H}{2}, the correlation is $<1\%$ with H$_2$. Similarly, the residuals of the fit with $\{$\ion{H}{1}-LVC, \ion{H}{1}-IVC, \ion{H}{2}$\}$ have a $\lesssim1\%$ level of correlation with the H$_2$ map. 

Therefore, we conclude that our H$_2$ template does not improve the modeling of dust emission in the high Galactic latitude regions considered in this work, and thus we do not incorporate the H$_2$ map in our subsequent analysis.

The lack of correlation between the H$_2$ template and the residuals from fitting \ion{H}{1} and \ion{H}{2} to the Planck map is driven by the noisiness of the CO map from which our H$_2$ template is built. The pixel values of the CO map in our regions are completely consistent with the noise map. H$_2$ may be present in non-negligible amounts along some of our high-latitude sight lines, either in the CO-dark phase or as a small fraction of H$_2$ traced by CO that falls below the sensitivity of our CO map. See Section~\ref{S:interp_gtdr} for further discussion.


\section{\texorpdfstring{Clustering-based \ion{H}{1} Model}{Clustering-Based HI Model}}\label{S:clustering_based_model}

In Section~\ref{S:multiphase}, we found that including H templates from non-neutral phases in the analysis modestly reduces the amplitude of the residuals compared to the \ion{H}{1}-only fit. Another potential avenue for improving the model is to construct a more sophisticated template set than a simple velocity cut for defining the \ion{H}{1} clouds, as has been used in previous work to define LVC and IVC components. In this section, we employ a data-driven technique to derive a set of \ion{H}{1} template maps from the 3D \ion{H}{1} data cube in the PPV space.

\subsection{Previous PPV-space Models}
The model presented in Equation~\eqref{E:linear_template_model} posits that all \ion{H}{1} gas within a discrete gas cloud has the same dust emissivity per H atom, while the dust emissivity generically differs from cloud to cloud. Since clouds are inherently localized in position and velocity phase space, ideally one would distinguish \ion{H}{1} clouds in a 6D velocity-spatial phase space. In practice, the line-of-sight distance and the transverse velocity information are not accessible from the \ion{H}{1} spectral images. Thus, we perform our analysis by separating these structures in PPV space into a few distinct clusters to serve as our \ion{H}{1} templates and determine the emissivity of each cluster by fitting to the Planck data. 

One example of this type of decomposition in PPV space is the work of \citet{2020ApJ...902..120P} that used Gaussian decomposition to construct an \ion{H}{1} cloud catalog from the HI4PI data. However, our focus is on much larger-scale coherent structures in \ion{H}{1}, and so we adopt a different algorithm as described below. A similar approach has also been employed in the analysis of CO emission in the Galactic plane by \citet{2017ApJ...834...57M}, where the CO emission is first decomposed into discrete Gaussian components. Then, neighboring components in PPV space are grouped into large clusters of structures. 

\citet{2019ApJ...883...75L} built a map of the CIB from Planck data by fitting the \ion{H}{1}-dust emission relation in a large number of sub-regions with $\sim$3.7$^\circ$ scales ($N_{\rm side}=16$) and several velocity bins. The effective number of velocity components used in each sub-region was determined by a regularization condition that reduced the number of degrees of freedom. However, coherent \ion{H}{1} structures can span several tens of degrees on sky and large velocity ranges, implying that independent fitting in small sub-regions and velocity bins may not effectively capture these features and could potentially lead to overfitting of the CIB signal into the Galactic emission. Section~\ref{S:cmblensing} further discusses this potential CIB signal loss in \citet{2019ApJ...883...75L}.

\subsection{Clustering Methodology}
We employ the unsupervised k-means clustering algorithm \citep{1056489} on the \ion{H}{1} PPV data cube to identify clusters of \ion{H}{1} structures. Since we target the large angular scale of \ion{H}{1} structures, we downsample the \ion{H}{1} maps in both spatial and velocity dimensions to reduce computational costs. We utilize the binned channel maps from \citet{2019ApJ...887..136C} that compressed the HI4PI maps at $|v|\lesssim 90$\,km\,s$^{-1}$ into 41 velocity bins (see Section~\ref{S:data}). When conducting the k-means clustering, we reduce the pixel resolution of the HI4PI data from its native $N_{\rm side}=1024$ to $N_{\rm side}=128$.

The k-means clustering algorithm classifies the input data into $k$ clusters that minimize the variance of the ``distance'' to the cluster center of all of the cluster members. The number of clusters, $k$, is left as a free parameter. When implementing the clustering algorithm, each voxel in PPV space is treated as a data point. Here, a voxel refers to a pixel in a single velocity bin map, which is the basis element of our dataset. For the attributes, we use the angular position\footnote{The angular position is converted to a Cartesian coordinate to define the distance metric. See Appendix~\ref{A:clustering} for details.}, velocity, and \ion{H}{1} channel density, defined as the \ion{H}{1} column density within the channel map divided by the channel velocity width ($\Delta N_{\rm H\text{\sc I}}^{\rm ch}/\Delta v^{\rm ch}$). For the distance metric, we use the Euclidean distance in the multidimensional space spanned by the data attributes described above. We impose weights on each attribute by scaling their contribution to the distance metric. These weights are considered as additional hyperparameters that we optimize in our fitting process. More details in the clustering algorithm implementation and the impact of different weightings on the clustering results is presented in Appendix~\ref{A:clustering}.

\begin{figure}
\begin{center}
\includegraphics[width=\linewidth]{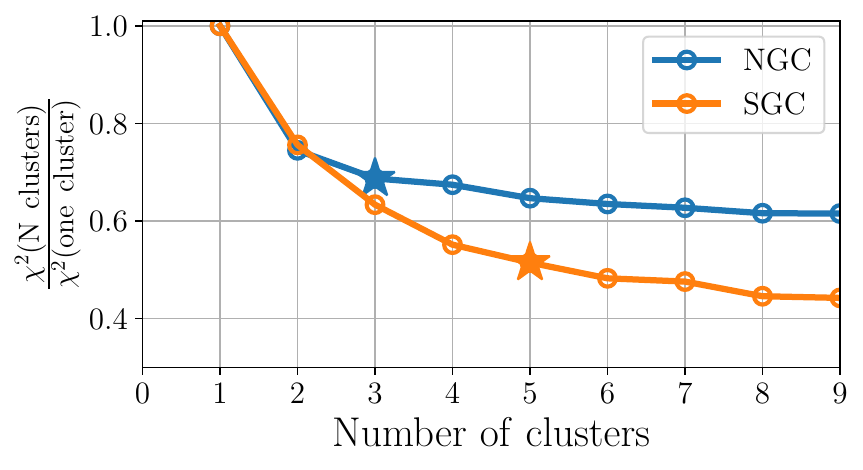}
\caption{\label{F:opt_Nclus} Minimum $\chi^2$ of fitting our model to the Planck 857\,GHz data in the NGC (blue) and SGC (orange) as a function of number of clusters $k$. For each $k$, we optimize for the attribute weights and the linear fitting coefficients. We choose three and five clusters for NGC and SGC, respectively, marked with the star symbols.}
\end{center}
\end{figure}

The NGC and the SGC regions are fitted separately. In each region, we iterate through different numbers of clusters ($k$). For each $k$, we search for the optimal weighting that minimizes the $\chi^2$ error between the Planck 857\,GHz map, $I^d_\nu$, and the model (Equation~\eqref{E:chi2}). We optimize based on fits at 857\,GHz since this band has the highest signal-to-noise ratio per pixel. For the model map (Equation~\eqref{E:HI_lin_fit}), we include only the \ion{H}{1} templates from our clustering algorithm and the \ion{H}{2} map, but not the H$_2$ component, since, as discussed in Section~\ref{S:interp_gtdr}, the H$_2$ has a negligible impact on the fit.

Figure~\ref{F:opt_Nclus} shows the minimum $\chi^2$ in 857\,GHz as a function of $k$ after optimizing the attribute weights and the linear fitting coefficients as described above. From Figure~\ref{F:opt_Nclus}, we see that the model performs better with increasing number of clusters $k$, but the improvement is saturated after a handful of clusters. This suggests that although we can identify more distinct structures in the \ion{H}{1} data, many of them have a similar emissivity ($\epsilon_{\nu,i}$ in Equation~\eqref{E:HI_lin_fit}), and thus the model cannot be further improved with even more clusters. Based on Figure~\ref{F:opt_Nclus}, we choose $k=3$ for the NGC and $k=5$ for the SGC in our model, which corresponds to the point where the $\chi^2$ improvement starts to saturate and where fits with additional clusters yield degenerate dust emissivities. 

\subsection{Results}\label{S:clus_results}

\begin{figure*}[!t]
\begin{center}
\includegraphics[width=0.48\linewidth]{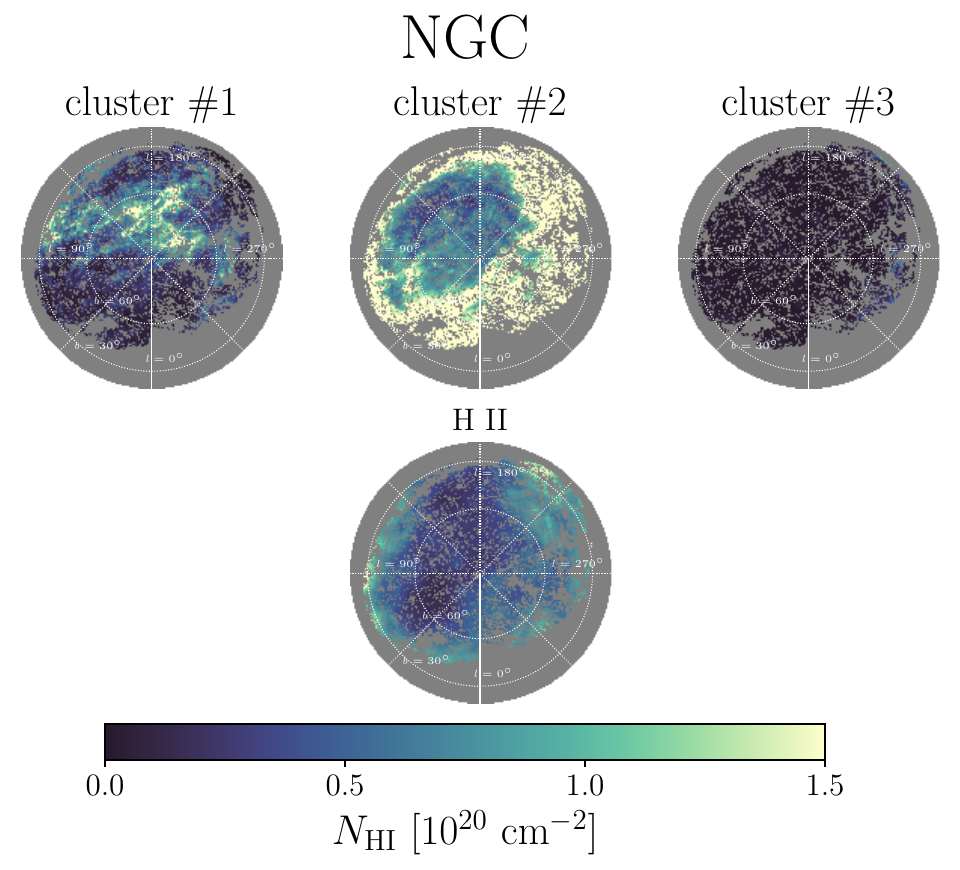}
\includegraphics[width=0.48\linewidth]{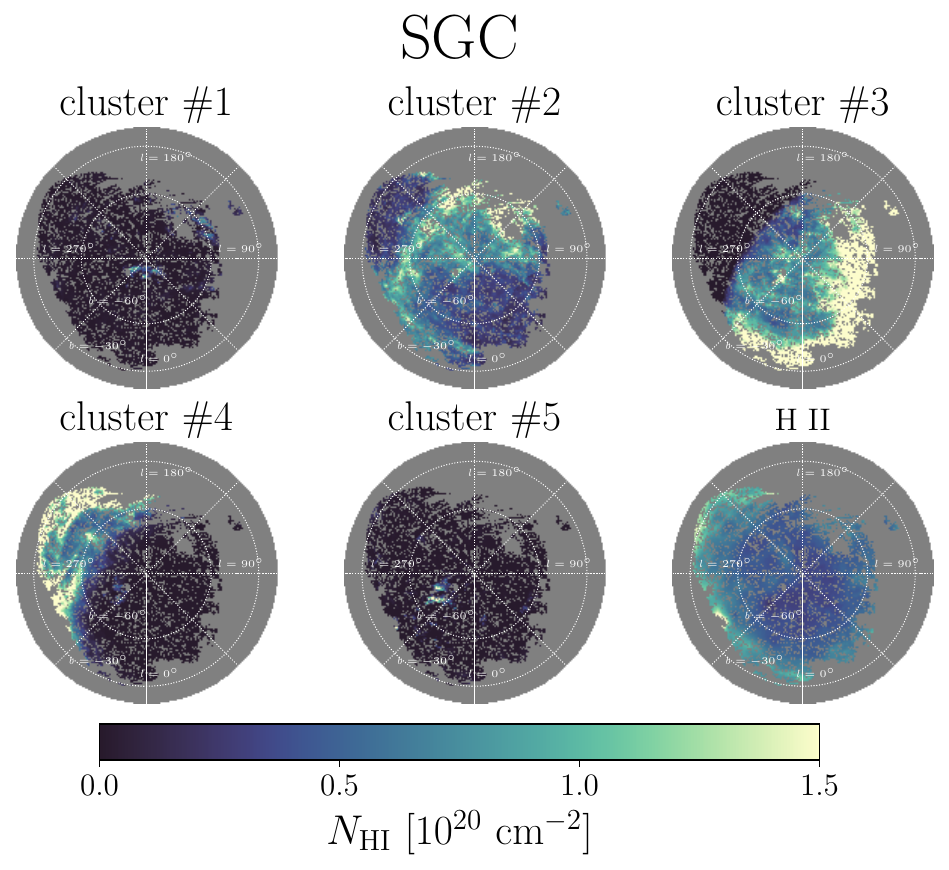}
\includegraphics[width=0.48\linewidth]{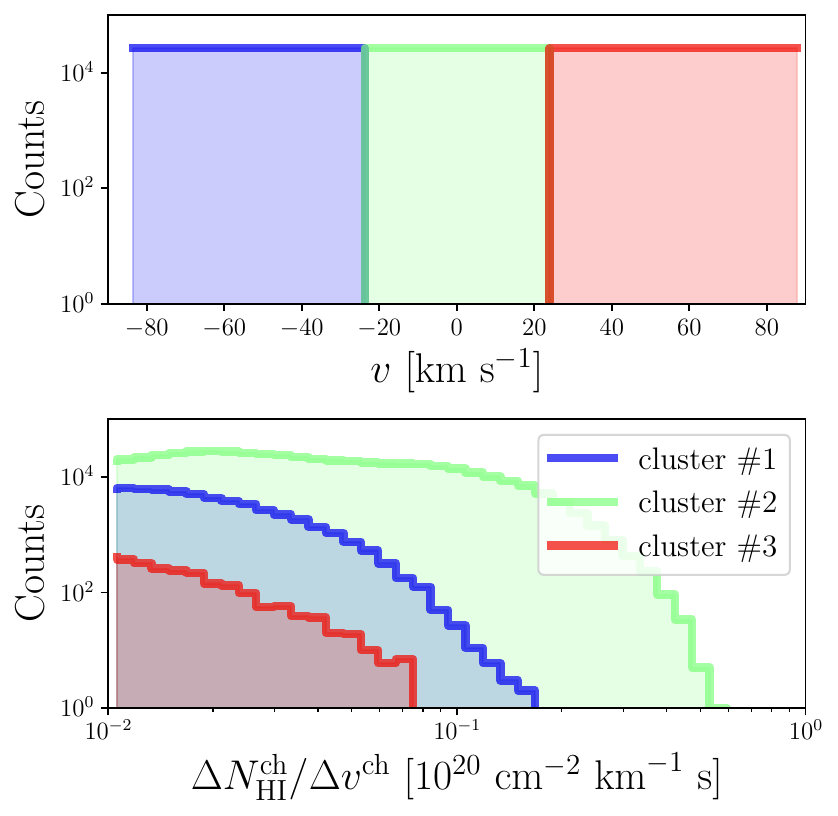}
\includegraphics[width=0.48\linewidth]{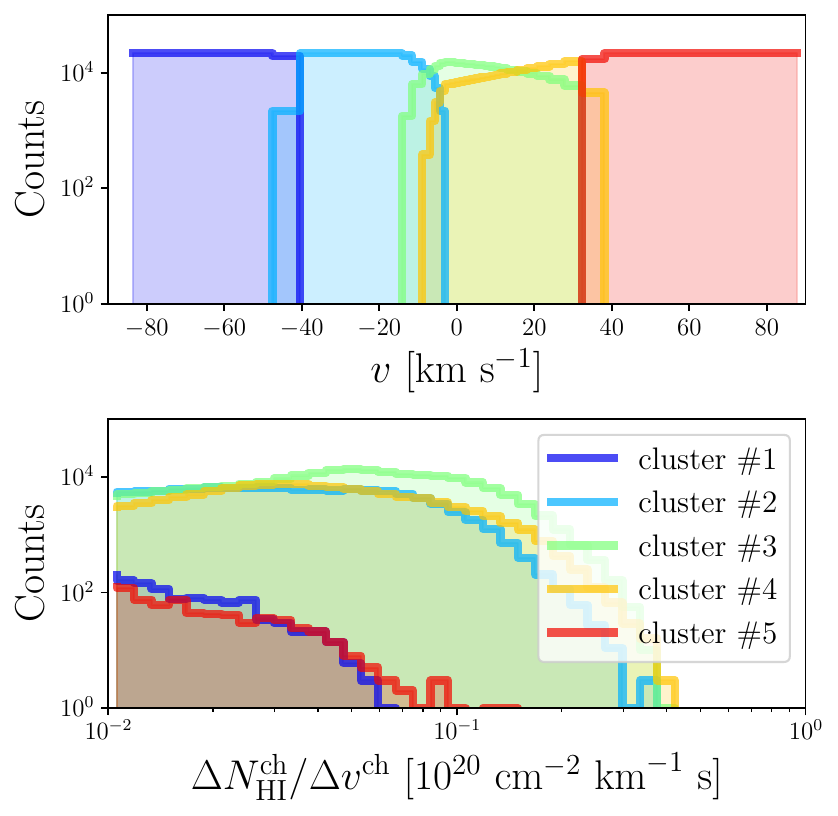}
\caption{\label{F:clus} Top left/right: the column density of the three/five \ion{H}{1} components in the NGC/SGC from our clustering algorithm and the \ion{H}{2} map. We denote cluster \#1 as ``IVC$-$'' and cluster \#2 as ``LVC'' in the NGC, and cluster \#2 as ``LVC$-$'' and the sum of clusters \#3 and \#4 as ``LVC$+$'' in the SGC. Middle left/right: histogram of the velocity of voxels in each cluster in the NGC/SGC. Bottom left/right: histogram of the \ion{H}{1} channel density of voxels in each cluster in the NGC/SGC.}
\end{center}
\end{figure*}

Figure~\ref{F:clus} displays the \ion{H}{1} column density maps and the histogram of the velocity and \ion{H}{1} density per voxel for the three/five clusters found from our k-means clustering algorithm in the NGC/SGC, which give the minimum $\chi^2$ value. 

In the NGC, while \ion{H}{1} structures span spatial, velocity, and \ion{H}{1} channel density domains, our results indicate that the best model for separating \ion{H}{1} components is a simple velocity cut, and $v=\pm 25$\,km\,s$^{-1}$ are the optimal velocity thresholds. This is very close to the previous Planck analysis \citep{planck2013-p06b} that defines the LVC and IVC with a $v=\pm 30$\,km\,s$^{-1}$ velocity cut. 

The IVC component with the negative velocity (cluster~\#1) corresponds to the location of the ``local chimney'' inferred from reconstructing the 3D structure of the Local Bubble \citep{2024ApJ...973..136O}. This suggests that the origin of this infalling IVC cloud could be part of the ISM that was expelled from the chimney and is now falling back to the Galactic disk \citep{1980ApJ...236..577B}. The IVC component with positive velocity (cluster~\#3) contains only a small fraction of the total \ion{H}{1} emission. When we perform a linear fit to the Planck 857\,GHz data with these three templates (Equation~\eqref{E:HI_lin_fit}), the coefficient $\epsilon_{\nu,i}$ of this template is negligible compared to the other two templates. Thus, we only take the clusters \#1 and \#2 as our \ion{H}{1} templates for the NGC. Hereafter, we name clusters \#1 and \#2 in the NGC as ``IVC$-$'' and ``LVC,'' respectively, based on the velocity ranges of the clusters.

For the SGC, our algorithm picks out two compact IVCs with high positive and negative velocities, respectively (\#1 and \#5). Similar to the NGC case, when we perform a linear fit to the Planck 857\,GHz data with these five templates (Equation~\eqref{E:HI_lin_fit}), the clusters~\#1 and \#5 have negligible coefficient $\epsilon_{\nu,i}$, and thus we only use the remaining three clusters for our templates. 

Clusters~\#3 and \#4 reside in the same velocity and \ion{H}{1} column density ranges and are therefore only separated based on their spatial distribution, where they occupy the $\ell\sim-60^\circ$---$180^\circ$ and $\ell\sim180^\circ$---$300^\circ$, respectively. When fitting with a four-component template set using clusters \#2, \#3, \#4, and \ion{H}{2} to Planck data, we find that cluster \#4 is highly degenerate with \ion{H}{2}, which also has high column density in the $\ell\sim225^\circ$ region. This results in the best-fit emissivity of \ion{H}{2} consistent with zero. On the other hand, if clusters \#3 and \#4 are combined into a single template, a three-component template fitting with this combined cluster, cluster \#2, and \ion{H}{2} yields $\epsilon_{\nu,{\rm HII}}$ to be about half of the $\epsilon_{\nu,{\rm HI}}$ fit to the clusters \#3$+$\#4 template (see Table~\ref{T:params}). While this $\epsilon_{\nu,{\rm HII}}$ value seems more physically reasonable, it is about twice as large as the $\epsilon_{\nu,{\rm HII}}$ value in the NGC. We choose to combine clusters \#3 and \#4 into a single template in our fiducial model. Therefore, in the SGC, we have two \ion{H}{1} templates: cluster \#2, hereafter referred to as ``LVC$-$'', and the combination of \#3 and \#4, hereafter referred to as ``LVC$+$''. More detailed discussion of the physical interpretations is provided in Section~\ref{S:combined_model}.

As in our analysis in Section~\ref{S:previous_model} based on simple \ion{H}{1} velocity cuts, we find that inclusion or exclusion of the H$_2$ map has a negligible impact on the results of these fits. This supports our decision not to incorporate the H$_2$ map into our fiducial template set.

Thus, to summarize, our fiducial template set consists of $\{$IVC$-$ (cluster \#1), LVC (cluster \#2), \ion{H}{2}$\}$ (NGC) and $\{$LVC$-$ (cluster \#2), LVC$+$ (cluster \#3 + cluster \#4), \ion{H}{2}$\}$ (SGC).

The residual map of fitting these fiducial templates to the Planck 857\,GHz map is shown in the bottom-right panel of Figure~\ref{F:clus_res_maps}. While visual inspection of the residual maps reveals that employing our new \ion{H}{1} templates does not result in a significant reduction of the residuals, compared to the case using the LVC and IVC defined by velocity cut (bottom-left panel of Figure~\ref{F:clus_res_maps}), our \ion{H}{1} templates improve the fit for the outlier sight lines, as illustrated by the reduction of the 3$\sigma$ interval of the residuals shown in Figure~\ref{F:clus_res_sigmas}.

\section{Combined Model}\label{S:combined_model}
\begin{figure*}[ht!]
\begin{center}
\includegraphics[width=0.32\linewidth]{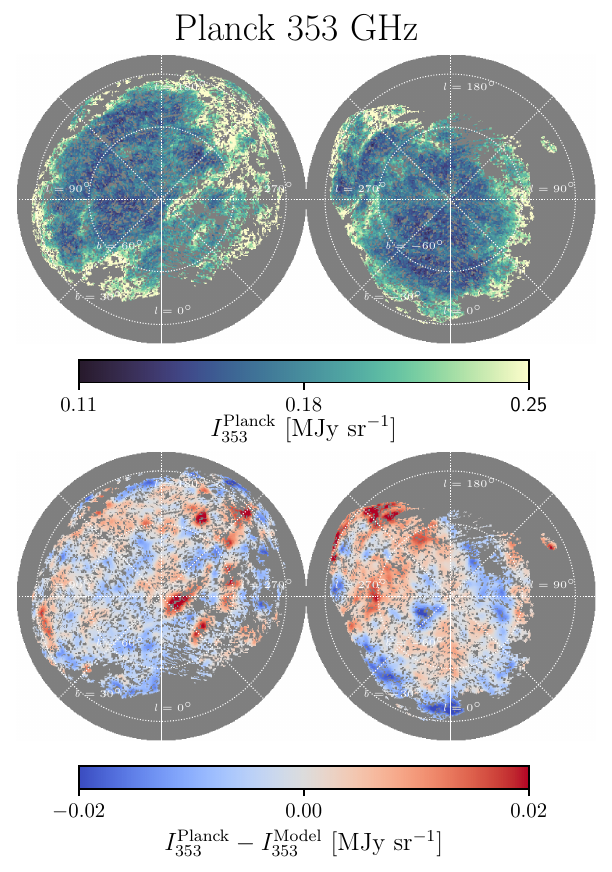}
\includegraphics[width=0.32\linewidth]{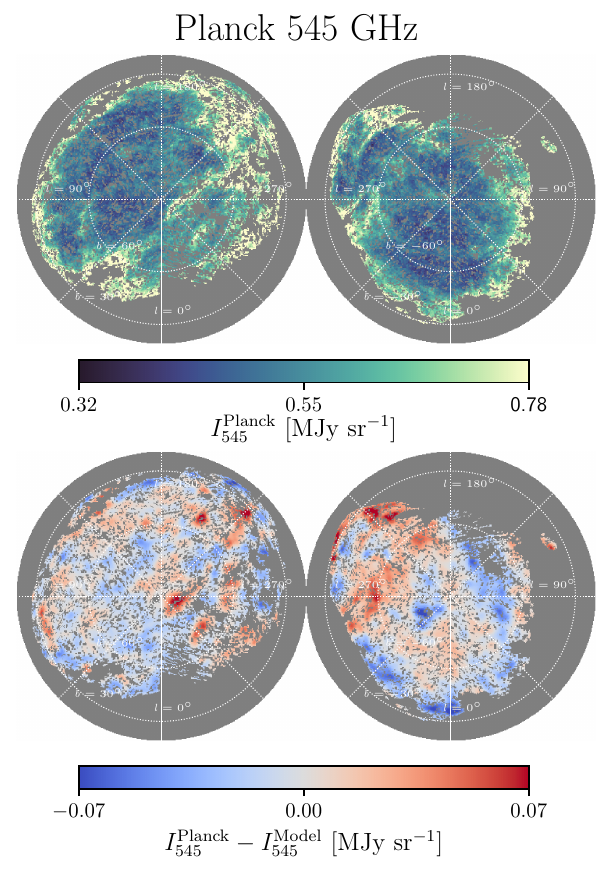}
\includegraphics[width=0.32\linewidth]{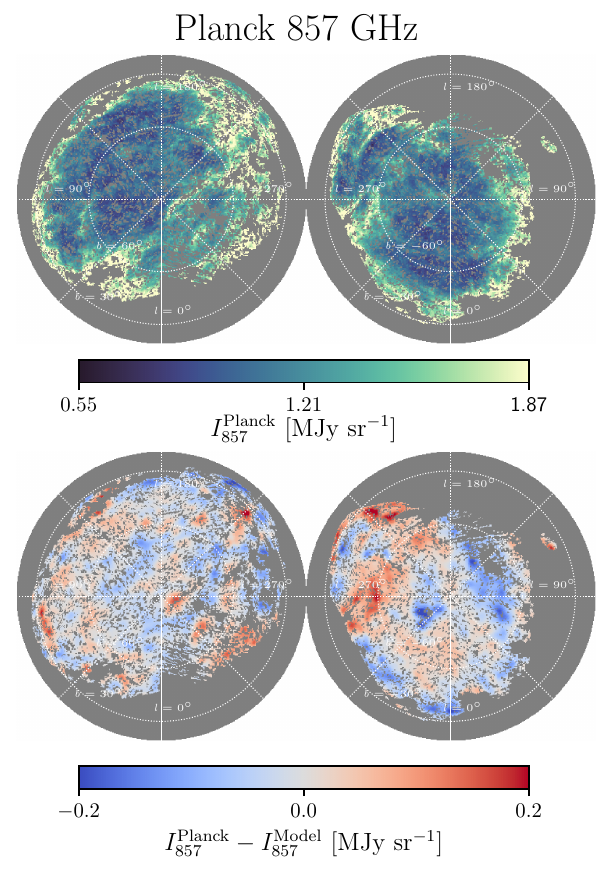}
\caption{\label{F:Planck_res} Top: Planck intensity maps in the 353 (left), 545 (middle), and 857 (right)\,GHz bands. Bottom: the residuals of fitting our multi-phase templates to the Planck map. We define the residual to be the data minus the model map $\delta I_\nu=I_\nu^d - I_\nu^m$. All residual maps are smoothed with a 1 degree Gaussian kernel to highlight the large-scale patterns.}
\end{center}
\end{figure*}

With our fiducial multiphase and clustering-based templates (Section~\ref{S:multiphase} and \ref{S:clustering_based_model}), we construct our final dust model through a linear combination of these templates described above. We then find the coefficients that minimize the $\chi^2$ upon fitting to the Planck maps (Equation~\eqref{E:chi2}) using the closed-form solution given by Equation~\eqref{E:ci_matrix}. This fitting is performed separately in the NGC and the SGC, and separately for each of the three Planck bands.

The best-fit emissivity values ($\epsilon_{\nu,i}$) from our model are summarized in Table~\ref{T:params}. Previous determinations of these parameters from Planck analyses \citep{planck2013-XVII} are also shown for comparison. The fitted emissivities of LVC in the NGC and LVC$+$ in the SGC show excellent consistency with each other and with those of \citet{planck2013-XVII}. The IVC$-$ in the NGC and the LVC$-$ in the SGC have the best-fit emissivities of $\sim$1.5 times lower than the LVC (NGC) and LVC$+$ (SGC), which is consistent with previous analyses that observed higher-velocity inflow of \ion{H}{1} to have lower dust emission \citep[e.g.,][]{planck2011-7.12, 2017ApJ...846...38L}.

However, the best-fit value of the \ion{H}{2} emissivity in the SGC is about twice that of the NGC. This discrepancy is likely due to the strong excess of dust emission at $\ell\sim225^\circ$ of the SGC, which cannot be fit by \ion{H}{1} alone. As our \ion{H}{2} template has higher column density in that region, the best-fit SGC emissivity is being driven to a higher value in order to reduce the residual in that region. As described in Section~\ref{S:multiphase}, the high \ion{H}{2} emissivity value in the SGC may be attributed to H$_2$ gas that is not directly traced by our template set but is correlated with the total column density, and therefore being absorbed in the \ion{H}{2} component fit. 

In addition, potential systematics may exist in the \ion{H}{2} template adopted from the DM map of \citet{2024A&A...690A.314H}. At high Galactic latitudes, the small-scale structures of the DM map are not well constrained due to the sparsity of their pulsar sample. The construction of their DM map also relies on H$\alpha$ EMs and free-free emission, both of which are subject to their own systematics and prior assumptions. Notably, \citet{2024A&A...690A.314H} found that their model underpredicts DM values at certain high-latitude sight lines when compared to previous measurements. Any systematics in the \ion{H}{2} template could introduce biases in our fitted coefficients and contribute to additional residuals in our model.

\begin{deluxetable}{c|ccc}[ht]
\tablenum{1}
\tablecaption{\label{T:params} Best-fit Emissivity from Our Model Compared to Previous Planck Analysis}
\tablewidth{0pt}
\tablehead{
\colhead{}  & \colhead{353 GHz} & \colhead{545 GHz} & \colhead{857 GHz}} 
\startdata
NGC $\epsilon_{\nu,{\rm IVC-}}$  & $0.026$ & $0.089$ & $0.254$  \\ 
NGC $\epsilon_{\nu,{\rm LVC}}$   & $0.038$ & $0.134$ & $0.392$  \\ 
NGC $\epsilon_{\nu,{\rm HII}}$   & $0.018$ & $0.052$ & $0.167$  \\ 
\hline
SGC $\epsilon_{\nu,{\rm LVC-}}$  & $0.025$ & $0.097$ & $0.306$  \\
SGC $\epsilon_{\nu,{\rm LVC+}}$  & $0.036$ & $0.127$ & $0.391$  \\
SGC $\epsilon_{\nu,{\rm HII}}$  & $0.040$ & $0.0137$ & $0.374$  \\
\hline
Planck $\epsilon_\nu$  & $0.039\pm0.0014$ & $0.14\pm0.015$ & $0.43\pm0.045$\\
\enddata
\tablecomments{Emissivity $\epsilon_\nu$ unit: MJy\,sr$^{-1}$\,($10^{20}$\,cm$^{-2}$)$^{-1}$; offset $b_\nu$ unit: MJy\,sr$^{-1}$. Planck $\epsilon_\nu$: emissivity from Planck analysis \citep{planck2013-XVII}.
}
\end{deluxetable}

Figure~\ref{F:Planck_res} shows the 353, 545, and 857\,GHz Planck intensity maps, as well as the residuals of our model fit defined by the difference of Planck map and our model map. Overall, our model can trace most of the emission observed by Planck, and the residuals are on the order of $\lesssim 20\%$ of the Planck intensity. The large-scale residuals exhibit a very similar pattern across frequencies. This suggests that these residuals are not caused by systematics in a single Planck map. 

While the pixel covariance is ignored in our model, a recent analysis has incorporated the pixel covariance from the CIB clustering \citep{2024MNRAS.531.4876A}. This analysis used a Bayesian framework to construct a spatially varying 353\,GHz emissivity map in the SGC using the Planck map and \ion{H}{1} data from the GASS survey \citep{2009ApJS..181..398M,2010A&A...521A..17K,2015A&A...578A..78K}. Despite the fact we use different \ion{H}{1} template maps and a different analysis method, our 353\,GHz residual map exhibits similar spatial patterns to their emissivity map. Thus, pixel-pixel covariance or details of our analysis choices are unlikely to account for the bulk of the residuals.


The variance of our fitting residuals is much larger than the measurement variance of the Planck map as well as the \ion{H}{1} and \ion{H}{2} template maps. Furthermore, there exist large-scale coherent structures in the residual maps, indicating that there are spatial variations in $\epsilon_{\nu}$ that have not been captured by our templates. Thus, the remaining residuals likely originate from mechanisms that cannot be captured by our multi-phase gas tracers. Therefore, the next objective of our analysis is to understand the origin of these variations.

\section{Interpretation}\label{S:interpretation}

\subsection{Overview}
The model developed in this work reproduces emission in three Planck FIR bands with residuals of $\lesssim 20\%$ over most of the high-latitude sky. However, the variance of the residual map exceeds the uncertainties of the template maps and of the Planck data, and the residuals display coherence over large angular scales, spanning tens of degrees.

\begin{figure}[ht!]
\begin{center}
\includegraphics[width=\linewidth]{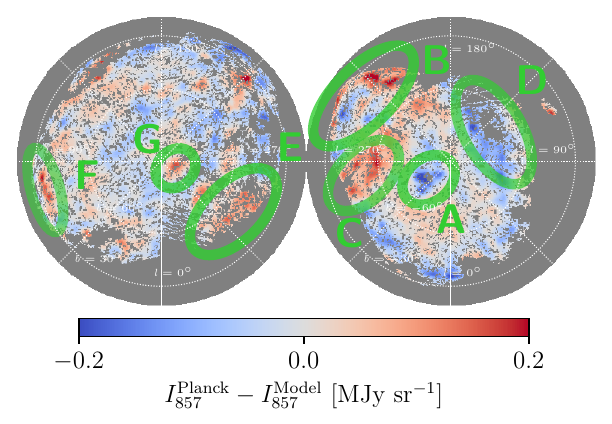}
\caption{\label{F:Planck_res_marked} Locations of Regions~A--G overlaid on the 857\,GHz residual map.}
\end{center}
\end{figure}

Sources of potential variation in the ratio of dust emission to $N_{\rm H}$ are evident in Equation~\eqref{eq:theory}: non-uniformity in the dust-to-gas ratio $\delta_{\rm DG}$, the dust temperature $T$, and/or the dust opacity law $\kappa_\nu$. It is also possible that despite using a multi-tracer approach for measuring $N_{\rm H}$, some gas remains unaccounted for. We examine each of these effects in this section, employing external datasets that trace different dust or gas properties to investigate the contributions from each. For ease of reference in the following discussion, we have marked a few regions with strong positive or negative residuals. The locations of these regions (referred to as Regions~A, B, C, D, E, F, and G) are shown in Figure~\ref{F:Planck_res_marked}, overlaid with the 857\,GHz residual map (same as the bottom right panel of Figure~\ref{F:Planck_res}).

\subsection{Comparison to Stellar Extinction Maps}
FIR dust emission and optical reddening both depend linearly on the dust column density. Thus, if $\delta_{\rm DG}$ variations were responsible for the majority of the residuals in our fits, we would observe a strong correlation between our residual map, produced by fitting our gas template maps to a Planck dust emission map, and the residuals obtained when performing the same fit of our gas template maps to a Galactic reddening map instead. We investigate this hypothesis using the stellar reddening maps built from the Dark Energy Spectroscopic Instrument (DESI) imaging and spectroscopy surveys \citep{2024arXiv240905140Z}. The stellar reddenings are determined by photometry in the $g$, $r$, and $z$ bands from the DESI Legacy Survey, with the intrinsic stellar parameters inferred from DESI spectra\footnote{We use their $E(B-V)$ map derived from $E(g-r)$ map on the pixel scale of $N_{\rm side}=128$. We have applied a multiplicative correction factor of $0.884$ from \citet{2011ApJ...737..103S} to their $E(B-V)$ map for consistency with our later analyses on the SFD map (see Section~\ref{S:data_release_sfd}).}.

\begin{figure}[!t]
\begin{center}
\includegraphics[width=\linewidth]{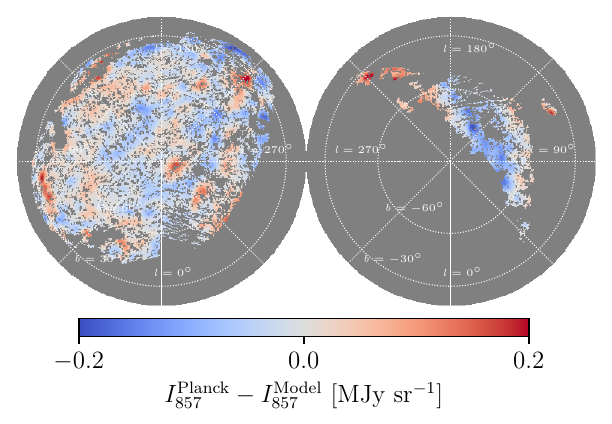}
\includegraphics[width=\linewidth]{desi_stellar_compare_desi.pdf}
\includegraphics[width=\linewidth]{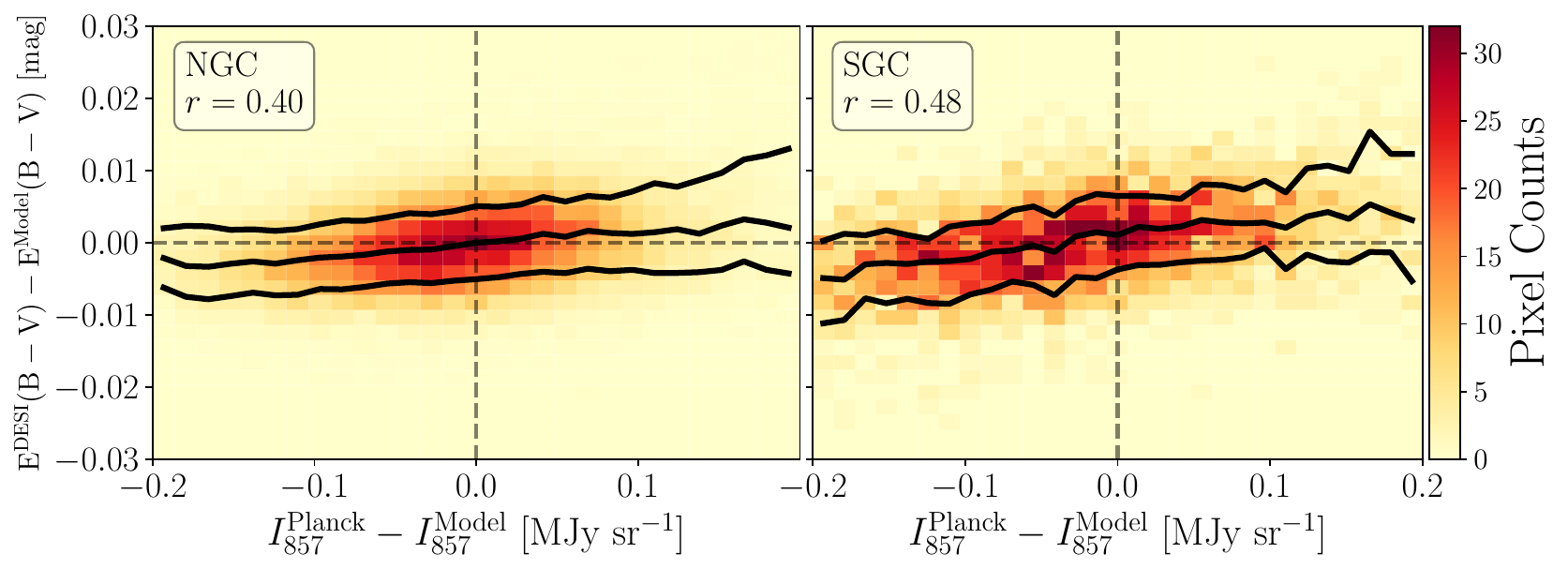}
\caption{\label{F:stellar_compare} Top: residual of fitting our multi-phase H templates to the Planck 857\,GHz map. Middle: residual of fitting the same multi-phase H templates to the stellar reddening map from \citet{2024arXiv240905140Z}. Both maps are smoothed with a $1^\circ$ Gaussian kernel to highlight the large-scale patterns. Bottom: the correlation between the 857\,GHz map residual and the stellar reddening map residual in the NGC (left) and SGC (right), respectively. Black solid lines mark the 16th, 50th, and 84th percentiles. The Pearson correlation coefficient $r$ is noted in the box.}
\end{center}
\end{figure}

Figure~\ref{F:stellar_compare} compares the residuals obtained from fitting our model to the Planck 857\,GHz map versus fitting to the stellar reddening map. The large-scale residuals of fitting our template to DESI and to the Planck map exhibit a correlation of $\sim$40--50\%. Moreover, we see a clear correspondence between the two residual maps in some regions, indicating that parts of these residuals are due to dust-to-gas ratio variations rather than systematics in the DESI reddening map. For example, Regions~B, D, and G have the same positive or negative excess in both residual maps. However, there are regions where the residuals do not exhibit the same spatial pattern. For instance, the positive residual in Region~E in the 857\,GHz fit has negative residual in the fit of the stellar reddening map. We conclude that it is difficult to explain the entirety of our fitting residuals residuals by dust-to-gas ratio variations.

We performed this analysis using several other stellar reddening maps from the literature \citep{2014ApJ...789...15S, 2019ApJ...887...93G, 2023ApJ...949...47M}. All of these reddening maps yielded residuals about twice as large as those of the DESI reddening map on degree scales, and thus we conclude that the DESI reddening map is less susceptible to large-scale systematics. The uncertainties of the DESI reddening map are about $3$\,mmag at half-degree resolution ($N_{\rm side}=128$). This is comparable to the residual fluctuation shown in Figure~\ref{F:stellar_compare}, suggesting that the statistical uncertainty of the DESI stellar reddening map has reached the level of the systematic error.

Large-scale variations of the dust-to-gas ratio have also been found in previous work using various observables tracing dust emission, extinction, or scattering. For instance, some previous analyses of the Planck data examined variations in the optical depth per $N_{\rm H}$ at 353\,GHz, $\tau_{\rm 353}/N_{\rm H}$. \citet{planck2013-p06b} produced a map of $\tau_{\rm 353}/N_{\rm H}$ by fitting an MBB model to the Planck 353, 545, and 857\,GHz bands and to the IRAS 100\,$\mu$m map. The total hydrogen content was derived from the sum of $N_{\rm HI}$ from the LAB survey \citep{2005A&A...440..775K} and $N_{\rm H_2}$ using a CO map from Planck \citep{planck2013-p03a, planck2014-a12}, assuming the same CO-to-H$_2$ conversion factor as employed in Section~\ref{S:H2_data}. Unlike our analysis, the \ion{H}{2} component is not included in their model of $N_{\rm H}$. Their findings revealed a nonuniform spatial distribution of $\tau_{\rm 353}/N_{\rm H}$, which shows correspondence with our residual maps in certain regions. For example, around Region~B where we observe a large positive residual, there is an excess of $\tau_{\rm 353}/N_{\rm H}$. This is not surprising as we find positive residuals in Region~B in all three Planck bands.

Measurements of dust FIR emission per H atom rely on an accurate estimate of the zero-point level in FIR maps. Previous determinations of the zero-point in Planck using correlations between intensity $I_\nu$ and \ion{H}{1} column density $N_{\rm HI}$ may be biased in regions containing a non-negligible fraction of ionized gas. To address this, \citet{2022ApJ...940..116C}  recalibrated the zero-point values of the 353, 545, and 857\,GHz Planck bands as well as the 100\,$\mu$m map from COBE/DIRBE and IRAS. They employed carefully selected sky regions where the gas column is dominated by \ion{H}{1}. After subtracting the recalibrated zero-point levels in these four frequency maps, they modeled the emission using an MBB model. With $N_{\rm HI}$ from the HI4PI data, they derived a map of $\tau_{\rm 353}/N_{\rm HI}$.

Their analysis revealed that the ratio of UV intensity to the \ion{H}{1} column density, $I_{\rm UV}/N_{\rm HI}$, exhibits correlated spatial structures with $\tau_{\rm 353}/N_{\rm HI}$. As UV emission arises from the scattering of stellar emission by Galactic dust, this finding presents additional independent evidence of the variation in the dust-to-gas ratio within the ISM. Their derived $\tau_{\rm 353}/N_{\rm HI}$ map shows similar excess as our residual maps in some regions such as the positive excess in Region~B. However, both the $\tau_{\rm 353}/N_{\rm H}$ map from \citet{planck2013-p06b} and the $\tau_{\rm 353}/N_{\rm HI}$ map from \citet{2022ApJ...940..116C} show only partial correspondence with our residual maps. This suggests that other factors likely also contribute to the residuals observed in our model.

Three-dimensional reddening maps have been produced using distance information from Gaia. To explore any distance-dependence in our residual maps, we employ the map\footnote{Data acquired from the \textit{dustmap} module \citep[][ \url{https://dustmaps.readthedocs.io/en/latest/modules.html}]{2018JOSS....3..695G}} of \citet{2019ApJ...887...93G}. We find that the majority of the signal in our residual map originates from the range of $\sim$0.2--0.3\,kpc.
\citet{2019A&A...631L..11S} also found that at high Galactic latitude, the polarized dust emission measured in Planck 353~GHz band is dominated by magnetized structures at the same range of line-of-sight distances, when jointly analyzed with polarized starlight data. 
Both of these results are in agreement with the fact that in the nearby region within $\lesssim 0.2$\,kpc, there is known to be an underdensity of dust and gas due to a supernova explosion \citep{2022Natur.601..334Z} and that most of the dust observed at high Galactic latitudes sits on the edge of this Local Bubble. 

An exception to this is that a significant portion of the positive residual in Region~B originates from distances of $\sim$0.3--$0.5$\,kpc. This indicates that variations in dust and/or gas properties across large physical scales in the Galaxy does indeed affect our fits even at fairly large angular scales. While this complexity may ultimately require 3D gas templates to address, it is evidently subdominant to variations in the more local ISM.

\subsection{Correlation with Gas Column Density}\label{S:interp_gtdr}
Gas and dust are well mixed in the ISM \citep[e.g.,][]{1996A&A...312..256B}, motivating our model of assigning a single emissivity to each gas component. However, our model is subject to two sources of uncertainty that produce similar signatures in residual maps. First, we may simply be underestimating the total gas column due to lack of sensitivity to H$_2$ and \ion{H}{2}. Second, there may be true astrophysical variations in the dust-to-gas ratio $\delta_{\rm DG}$. We investigate each of these in turn.

\begin{figure}
\begin{center}
\includegraphics[width=\linewidth]{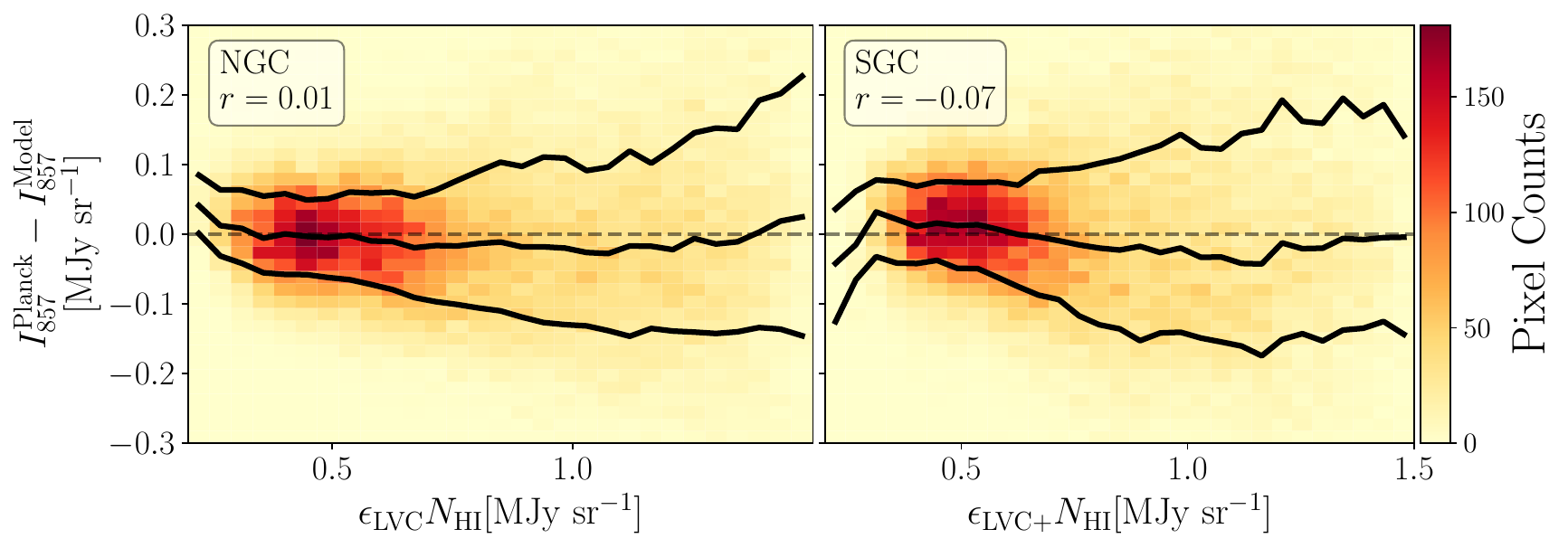}
\includegraphics[width=\linewidth]{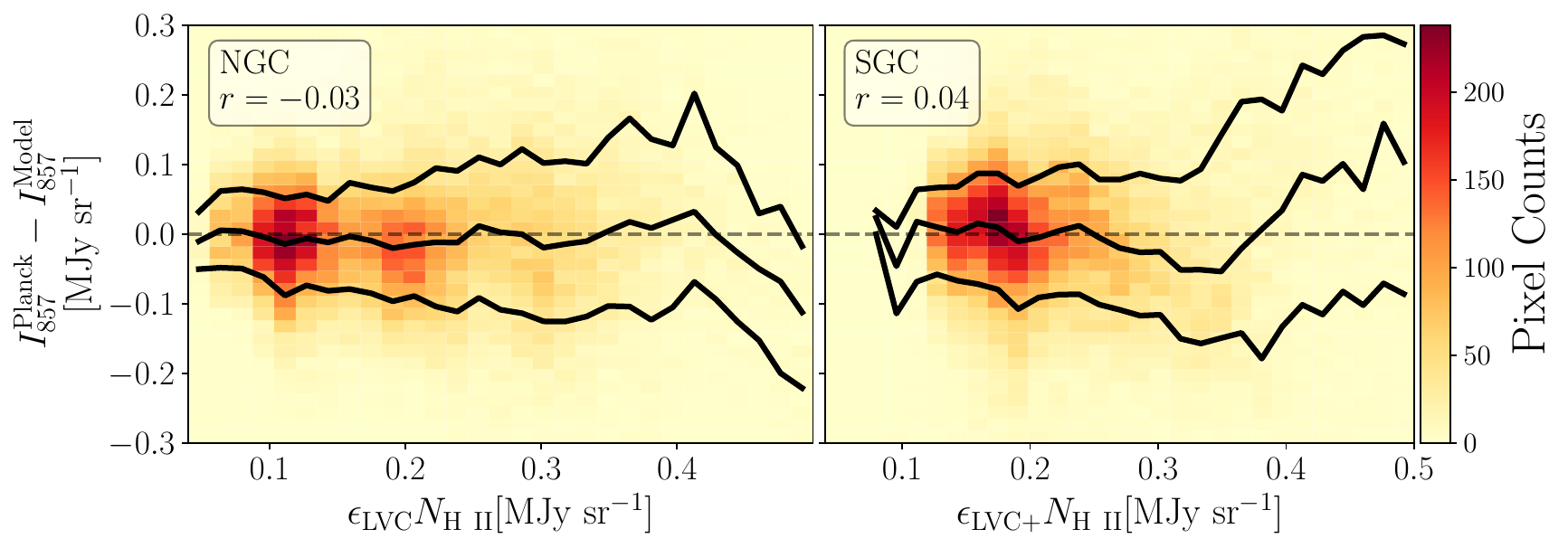}
\caption{\label{F:res_H_hist} Correlation between the 857\,GHz residual map and the \ion{H}{1} (top) and \ion{H}{2} (bottom) column density maps in the NGC (left) and SGC (right), multiplied by $\epsilon_{\rm LVC}$ (NGC) or $\epsilon_{\rm LVC+}$ (SGC). Black solid lines mark the 16th, 50th, and 84th percentiles. The Pearson correlation coefficient $r$ is noted in the box.}
\end{center}
\end{figure}

To test for possible unaccounted gas in our model, particularly missing H$_2$ or \ion{H}{2}, we first examine the correlation of our residuals with the \ion{H}{1} and \ion{H}{2} column densities, as shown in Figure~\ref{F:res_H_hist}. In both cases, we find no evidence of correlation, either positive or negative, between the fit residuals and the \ion{H}{1} or \ion{H}{2} column density. This suggests that our model has already extracted all of the information that can be gleaned from these gas tracers at large angular scales.

H$_2$ is expected to trace higher column density sight lines where self-shielding is sufficient. At high Galactic latitude, the self-shielding transition occurs at $N_{\rm H} \sim 2.5 \times 10^{20}$ cm$^{-2}$, with the molecular fraction ranging from $1\%$ to $30\%$ above this threshold \citep{2006ASPC..348..439G}. Given that our mask selects sight lines below $N_{\rm H} = 4 \times 10^{20}$ cm$^{-2}$, it is expected that some of our higher column density regions contain a non-negligible amount of H$_2$. However, as discussed in Section~\ref{S:multiphase}, part of the dust emission associated with H$_2$ is likely fitted by our \ion{H}{1} and/or \ion{H}{2} templates, as all of these components positively correlate with the total column density. Nonetheless, an excess of positive residuals in high $N_{\rm HI}$ regions could indicate that missing H$_2$---not perfectly accounted for by our existing templates---contributes to the overall residuals. Since this effect is not observed in Figure~\ref{F:res_H_hist}, it suggests that the lack of H$_2$ tracers is not a major source of residuals.

\begin{figure}[ht!]
\begin{center}
\includegraphics[width=\linewidth]{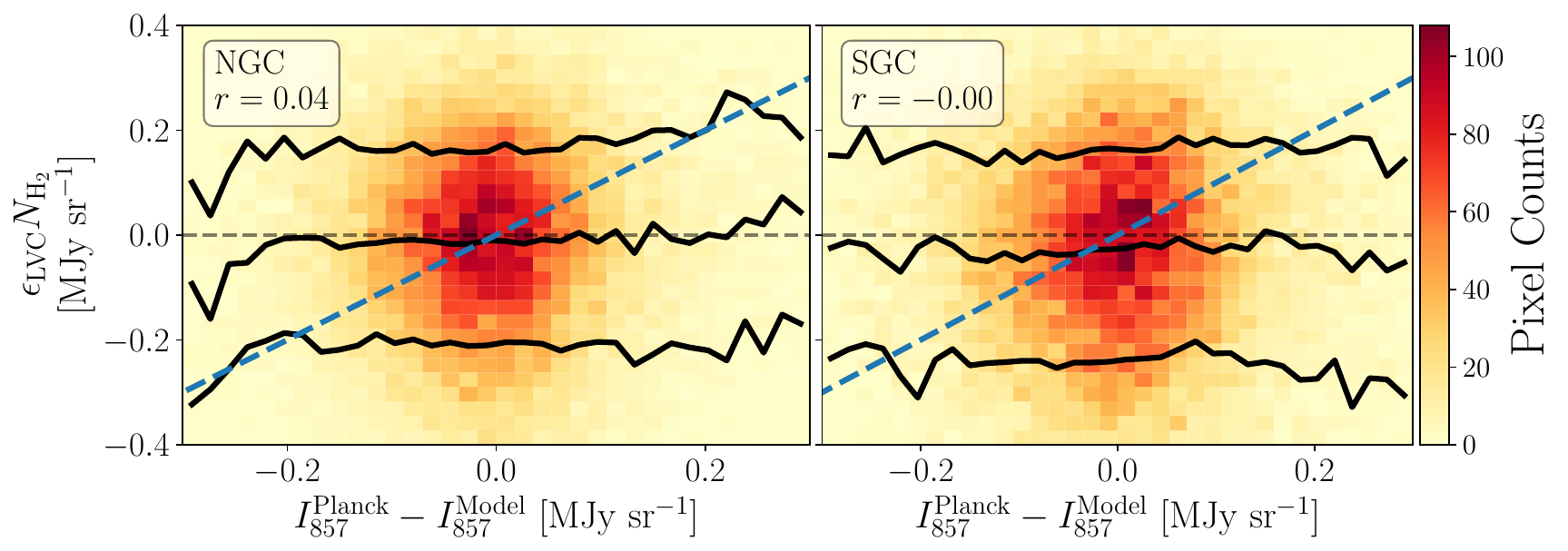}
\caption{\label{F:res_H2_hist} Correlation between the 857\,GHz fit residuals and the H$_2$ column density in the NGC (left) and SGC (right), multiplied by $\epsilon_{\rm LVC}$ (NGC) or $\epsilon_{\rm LVC+}$ (SGC). Black solid lines mark the 16th, 50th, and 84th percentiles. Blue dashed lines are a unity slope line for reference. Residual H not captured by our model would exhibit a positive slope and would be parallel to the blue dashed line if the residual hydrogen gas has the same emissivity as the LVC (NGC) or LVC+ (SGC). The Pearson correlation coefficient $r$ is noted in the box.}
\end{center}
\end{figure}

While we find no correlation between our residuals and the \ion{H}{1} column density, molecular gas merits additional consideration given the role that the \ion{H}{1}-FIR and \ion{H}{1}-$E(B-V)$ connections have played in inferring the presence of molecular gas \citep[e.g.,][]{Heiles:1988, planck2011-7.0, Kalberla:2020, Skalidis:2024}. We use the CO map as a proxy for the H$_2$ gas density. However, as described in Section~\ref{S:multiphase}, we found that adding the H$_2$ template to the fit does not reduce the residuals, and thus we have omitted the H$_2$ template in our fiducial model. The CO signal map in our high-latitude regions is completely consistent with the noise map. To search for a statistical detection of molecular gas, Figure~\ref{F:res_H2_hist} examines the amplitude of the residuals as a function of the CO intensity. There is no correlation. If unaccounted H$_2$ with associated CO emission has the same dust emissivity per H atom as we find in the LVC (NGC) / LVC+ (SGC) templates, we would have observed a positive correlation following the blue dashed line in Figure~\ref{F:res_H2_hist}. It is unclear whether the needlet-based analysis of \citet{2024A&A...688A..54G} would have been sensitive to so small a CO signal through a statistical average over many pixels, or whether any CO signal below a certain sensitivity threshold simply would not impact the final maps. Future observations with higher-sensitivity CO mapping in high-latitude regions are needed to further constrain the contribution of molecular components in the model.


In the low column density regime ($N_{\rm HI} \lesssim 4 \times 10^{20}$\,cm$^2$), the majority of the gas is in the neutral phase, and \ion{H}{1} has been shown to exhibit a strong linear correlation with both the extinction and emission of the Galactic dust \citep{planck2013-p06b, 2017ApJ...846...38L}. However, exceptions arise for inflowing gas, which is generally of low metallicity and thus has a lower dust emissivity per H atom \citep{Wakker:1997}. Inflowing gas is typically distinguished from gas of the diffuse ISM of the Galaxy by its large negative velocity. Our results are in agreement with this picture, as we find emissivities associated with negative velocity components are lower than the LVC (see Table~\ref{T:params}).

It is plausible that our clustering algorithm fails to identify some of these low-metallicity clouds, particularly if they extend to intermediate or low velocities. Clustering algorithms are well suited for segregating clusters of data points that are concentrated in parameter space and clearly separated from the rest of the dataset. However, \ion{H}{1} gas is known for its filamentary structure \citep{2006ApJ...652.1339M,2014ApJ...789...82C,2020A&A...642A.163S,2023ApJ...946..106C, 2023MNRAS.tmp.2686K, 2024ApJ...961...29H}, which can connect different cloud clusters through filaments, potentially undermining our algorithm's ability to separate components effectively. The fact that there is large-scale structure in the residual maps spanning tens of degrees on the sky suggests instead that it is insufficient to model variations in dust and/or gas properties in terms of discrete clouds with this algorithm.

Regions with low-metallicity gas not captured by our clustering algorithm would produce negative residuals (i.e., less dust emission than expected from $N_{\rm HI}$). Our clustering analysis identified two IVCs in the SGC in Region~A that exhibit lower dust emission to gas ratios compared to the rest of the \ion{H}{1} gas (clusters \#1 and \#5 in the SGC in Figure~\ref{F:clus}). Even after removing these two clusters, our model still displays a slightly negative residual in Region~A. This area in the sky corresponds to the Magellanic Stream \citep{2010ApJ...723.1618N}. Our clustering algorithm has effectively identified both positive and negative IVC components of the Magellanic Stream. However, it is expected that between the transitions of these two high-velocity components, low-velocity \ion{H}{1} gas associated with the Magellanic Stream exists but that is not being captured by our clustering algorithm. As a result, we attribute the negative residual in Region~A to the remaining Magellanic Stream gas within our \ion{H}{1} template.

The positive residuals in Region~C correspond to the locations of the Large and Small Magellanic Clouds, and the Region~G is the location of Markkanen’s cloud \citep{1979A&A....74..201M}. These clouds have different dust-to-gas properties compared to the Milky Way average. Although large parts of these clouds have been masked by our \ion{H}{1} density threshold at $4\times10^{20}$ cm$^{-2}$, there are still features in the residual map in their outskirt regions. Furthermore, much of the gas in these clouds is at high radial velocity. Since we use only \ion{H}{1} data with velocities within $|v|<90$\,km\,s$^{-1}$, the dust associated with these higher-velocity components is not accounted for in our model. This also contributes to the underestimation of the dust emission at these locations in our model.

Aside from these regions, we do not find significant overlap between our negative residuals and any HVC structures identified in a dedicated search by \citet{2018MNRAS.474..289W}. Unless there is a significant reservoir of low-metallicity gas in the LVC/IVC velocity range without a corresponding HVC component, it is unlikely that such gas is a dominant contributor to our residuals.

\begin{figure}[ht!]
\begin{center}
\includegraphics[width=\linewidth]{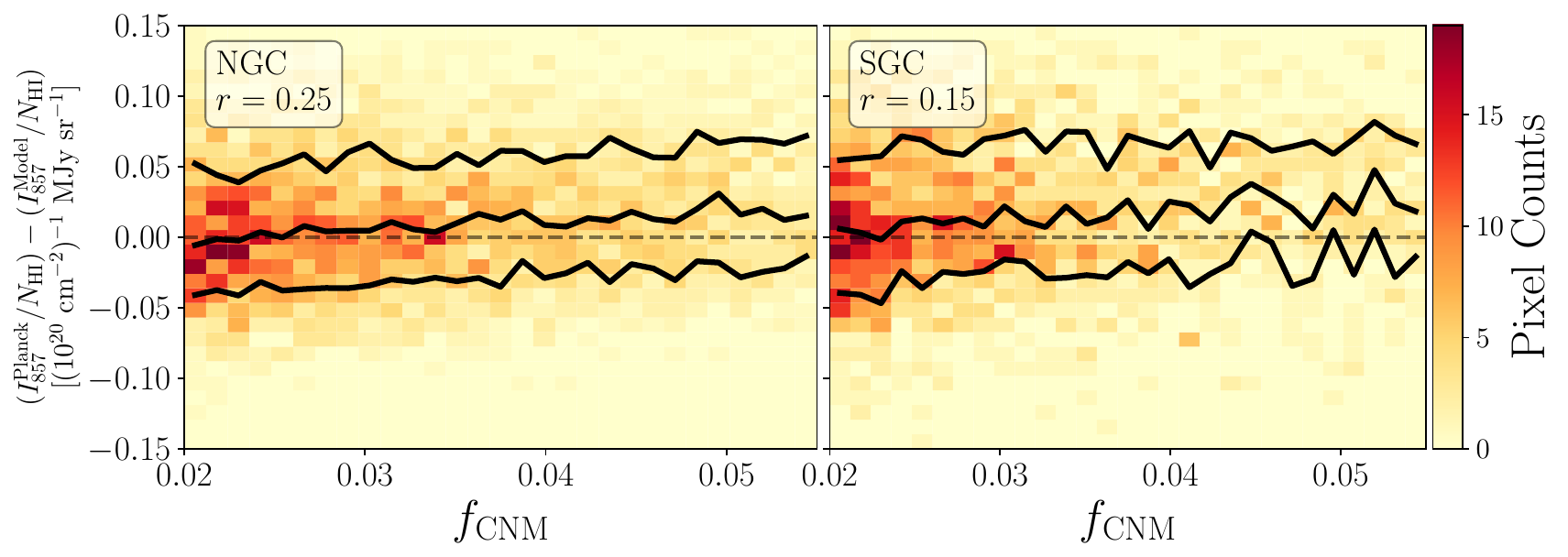}
\caption{\label{F:fcnm_compare} Correlation between the 857\,GHz residual per \ion{H}{1} column density and the CNM fraction $f_{\rm CNM}$ in the NGC (left) and SGC (right). Black solid lines mark the 16th, 50th, and 84th percentiles. The Pearson correlation coefficient $r$ is noted in the box.}
\end{center}
\end{figure}

The dust-to-gas ratio may be systematically higher in relatively cold, dense gas. \citet{2020ApJ...899...15M} found that the ratio of Planck 857\,GHz intensity to the \ion{H}{1} column density, $I_{857}/N_{\rm HI}$, increases with the fraction of cold neutral medium (CNM) in the ISM ($f_{\rm CNM}$). We also find a slight positive correlation between our residual map in the 857\,GHz band and the $f_{\rm CNM}$ map from \citet{2022ApJ...929...23H}\footnote{\url{https://dataverse.harvard.edu/dataset.xhtml?persistentId=doi:10.7910/DVN/72VS6T}}, as shown in Figure~\ref{F:fcnm_compare}. The $f_{\rm CNM}$ map is derived from applying the convolutional neural network developed by \citet{2020ApJ...899...15M} to the HI4PI data. As we find our residuals are not correlated with H$_2$, this slight correlation could point to the presence of molecular gas below current CO detection limits, or possibly an enhanced dust to gas ratio in the CNM phase. However, the weakness of the correlation in Figure~\ref{F:fcnm_compare} prevents us from placing meaningful constraints on the emissivity of the CNM phase.

\subsection{Temperature Variation}\label{S:T_var}
In addition to variation in dust-to-gas ratio, other potential sources of the fitting residual could be the variations in dust temperature $T$ and/or the opacity law, often modeled as a power law $\kappa_\nu \propto \nu^\beta$. Consider the MBB model (Equation~\eqref{E:MBB}), which can be written as:
\begin{equation}
\epsilon_\nu \propto \frac{\nu^{3+\beta}}{e^{h\nu/k_B T}-1}.
\end{equation}
Unlike variations in the dust-to-gas ratio, which cause the emissivity $\epsilon_\nu$ to change by the same multiplicative factor in all bands, changes in $T$ and/or $\beta$ result in a frequency-dependent modification of $\epsilon_\nu$. Therefore, the effects of dust-to-gas ratio variations and those of $T$ and $\beta$ can be distinguished by examining the frequency dependence of the residuals.

Previous studies have found evidence of variations in the spectral slope of $\epsilon_\nu$ in Planck data. For example, \citet{2024ApJ...970...43S} explored the spatially varying spectral slope of dust emission at high Galactic latitudes using Planck maps spanning from 100 to 857\,GHz. Employing a Gaussian mixture model with Bayesian likelihood analysis, they found that the data favor a two-component dust emission model with distinct spectral slopes corresponding to a slight difference in dust temperature. This result indicates that Galactic dust properties exhibit spatial variation, which may account for some of our residuals when fitting the Planck map with our gas templates.

In this section, we discuss the possibility that our fitting residuals are the result of temperature variation. Throughout, we assume a constant $\beta = 1.5$. The variation of the opacity law will be discussed in Section~\ref{S:beta_var}. We assume no variation in the dust-to-gas ratio $\delta_{\rm DG}$ and no unaccounted gas components from our template in order to quantify the upper bound of the temperature variation contribution to the residuals. We also assume a constant temperature along each line of sight. This assumption is adopted in most of the previous analyses \citep[e.g.,][]{planck2014-a12, planck2016-XLVIII}.

If all residuals are caused by temperature variation, one can derive a temperature map from the mismatch between the observed data and the model. With a spatially varying temperature $T(\hat{\theta})$, the observed intensity map can be expressed as
\begin{equation}
I_\nu^d(\hat{\theta}) = \left[\sum_{i}^{N_c} \epsilon_{\nu, i}(T(\hat{\theta})) N_{{\rm H}, i}(\hat{\theta})\right] + b_\nu,
\end{equation}
where the first term in brackets represents the total intensity from different gas components (including gas in all phases). Here, we assume the same temperature for all gas components along a given line of sight $\hat{\theta}$. Our modeled intensity, derived by fitting a linear model to the data, is given by
\begin{equation}
I_\nu^m(\hat{\theta}) = \left[\sum_{i}^{N_c} \epsilon_{\nu, i}(\overline{T}) N_{{\rm H}, i}(\hat{\theta})\right] + b_\nu,
\end{equation}
where we assume a constant mean temperature $\overline{T}$ for all components. If temperature fluctuations are the only source of residual error, i.e., there are no spatial fluctuations in the dust-to-gas ratio or opacity, the ratio of the true and modeled emissivities is given by
\begin{equation}
\frac{\epsilon_{\nu, i}(T(\hat{\theta}))}{\epsilon_{\nu, i}(\overline{T})}=\frac{e^{h\nu/k_B\overline{T}}-1}{e^{h\nu/k_BT(\hat{\theta})}-1} = \frac{I_\nu^d(\hat{\theta})-b_\nu}{I_\nu^m(\hat{\theta})-b_\nu}.
\end{equation}
Using this relation, we can derive a temperature map from the difference between the measured and modeled intensity maps, which can be expressed as
\begin{equation}
T(\hat{\theta}) = \overline{T}\frac{h\nu}{k_B\overline{T}}\left\{{\rm log}\left[\frac{I_\nu^m(\hat{\theta})-b_\nu}{I_\nu^d(\hat{\theta})-b_\nu}\left(e^{h\nu/k_B\overline{T}}-1\right)\right]\right\}^{-1}.
\end{equation}

\begin{figure}
\begin{center}
\includegraphics[width=\linewidth]{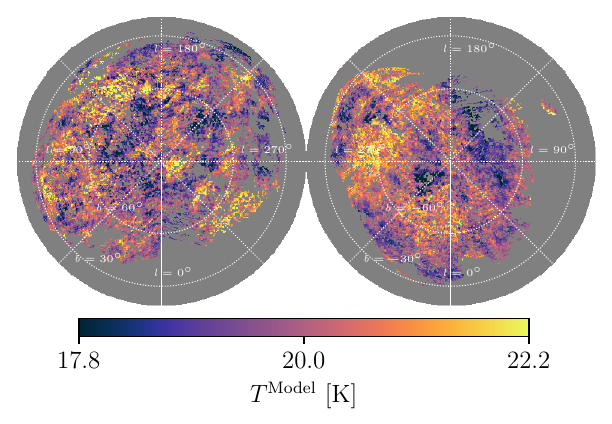}
\includegraphics[width=\linewidth]
{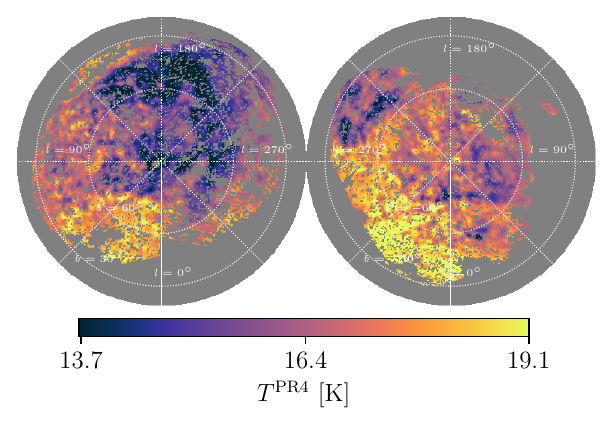}
\caption{\label{F:T_BP_map} Top panel: inferred temperature variation derived under the assumption that the fitting residuals at 857~GHz are entirely attributed to temperature fluctuations. Bottom panel: dust temperature map from the Planck PR4 \texttt{NPIPE} analysis.}
\end{center}
\end{figure}

Assuming a mean temperature of $\overline{T}=20$~K, and using the Planck map and our model at 857~GHz, we derived the implied temperature map, shown in the top panel of Figure~\ref{F:T_BP_map}. Our derived temperature variation map has a standard deviation of $\sigma_T=1.02$~K. If we consider a conservative upper limit of $\overline{T}<25$~K, we get $\sigma_T<1.28$~K. As this estimation assumes the residuals are entirely due to temperature variation, our derived limit, $\sigma_T<1.28$~K, sets an upper bound on the possible temperature variation in our high-latitude fields. 

We note that this temperature variation limit is derived under the assumption of a constant temperature along each line of sight; a higher temperature variation is plausible if there are multiple dust components with different temperatures in a given sight line. Some previous Planck analyses that also assume a constant line-of-sight temperature derive the high-latitude dust temperature variation in the range of $\sim$1.5--2.5~K \citep{planck2016-XLVIII}, higher than our upper limit, suggesting that their temperature maps are subject to systematic errors. One possible reason for the higher temperature variation found in previous Planck analyses is that they fit for both $T$ and $\beta$ in their model. The strong covariance between these two parameters in the MBB model can contribute to a higher derived temperature variation.

We compare our derived temperature map to those from SFD \citep{1998ApJ...500..525S}\footnote{\url{https://dataverse.harvard.edu/dataset.xhtml?persistentId=doi:10.7910/DVN/EWCNL5}} and from Planck analyses with \texttt{Commander} \citep{planck2014-a12}\footnote{\texttt{COM\_CompMap\_dust-commander\_0256\_R2.00.fits}} and GNILC \citep{planck2016-XLVIII}\footnote{\texttt{COM\_CompMap\_Dust-GNILC-Model-Temperature\_2048\_R2.01.fits}} methods of component separation. However, none of these temperature maps exhibits a positive correlation with our residual maps at a $>5\%$ level. Furthermore, these three temperature maps display significant inconsistencies with each other, indicating that the derivation of these temperature maps is highly susceptible to systematics or parameter degeneracy. Therefore, they cannot be considered reliable datasets for inferring temperature fluctuations in our analysis. 

The latest Planck PR4 uses the \texttt{NPIPE} analysis method to jointly process the temperature and polarization data from both Planck low- and high-frequency instruments \citep[LFI/HFI;][]{2020A&A...643A..42P}. The resulting mean value of dust temperature map from their component separation, as released by \citet{planck2020-LVII}\footnote{\url{http://sdc.uio.no/vol/cosmoglobe-data/BeyondPlanck/releases/v2/BP_dust_IQU_n1024_v2.fits}}, is shown in the bottom panel of Figure~\ref{F:T_BP_map}. The temperature variation from the Planck PR4 map over this region of sky is $\sim$1.5~K, slightly higher than our limit of $\sigma_T<1.28$~K. Furthermore, the large-scale patterns of the temperature maps are not well matched to each other, except for a few regions, such as the lower temperature observed in Region~D. No negative residuals in Region~D appear in our fit to the stellar extinction map (see Figure~\ref{F:stellar_compare}). Thus, dust temperature variation might partly contribute to the emission deficit observed in Region~D.

\begin{figure}[ht!]
\begin{center}
\includegraphics[width=\linewidth]{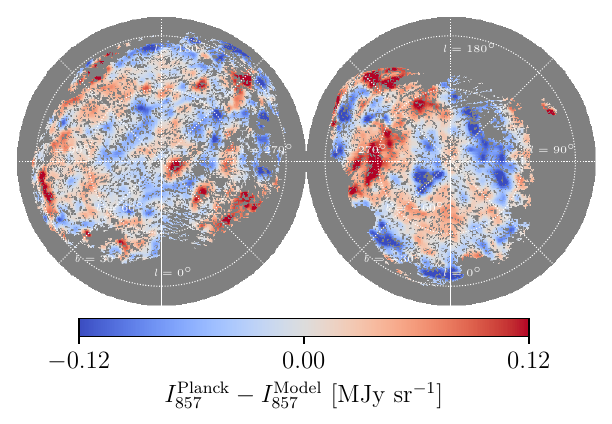}
\includegraphics[width=\linewidth]{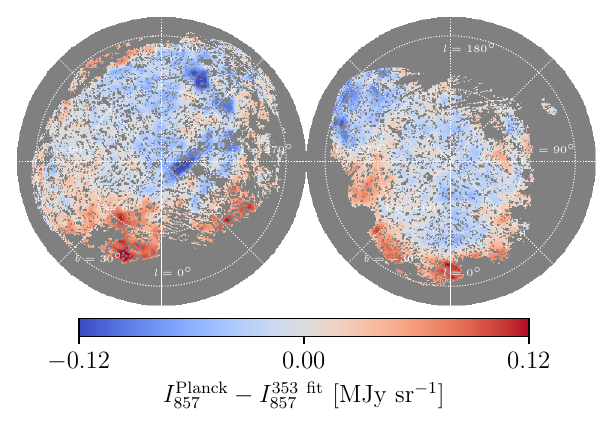}
\includegraphics[width=\linewidth]{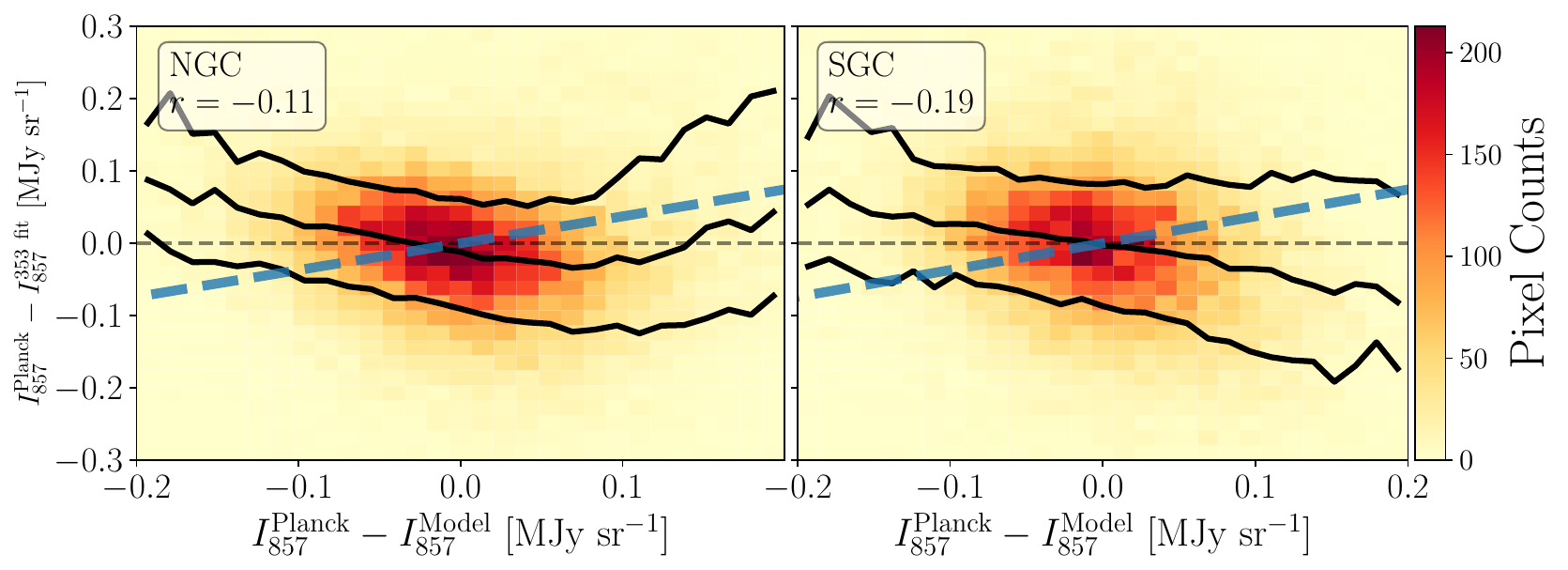}
\caption{\label{F:353_compare} Top: residual of fitting our multi-phase H templates to the Planck 857\,GHz map. Middle: residual of fitting the Planck 353\,GHz map to the 857\,GHz map. Both maps are smoothed with a $1^\circ$ Gaussian kernel to highlight the large-scale patterns. Bottom: the correlation between the 857\,GHz map residual fitted with our \ion{H}{1} templates and with the Planck 353\,GHz map in the NGC (left) and SGC (right). Black solid lines mark the 16th, 50th, and 84th percentiles. The Pearson correlation coefficient $r$ is noted in the box.}
\end{center}
\end{figure}

Another test for dust temperature variations is examining the ratio of residual maps across different frequencies. Dust-to-gas ratio variations affect each band through the same multiplicative factor, resulting in a uniform ratio of residual maps, while temperature variations do not. We employed this test by taking the Planck 353\,GHz map as a single-component template and fitting a linear model to the 857\,GHz map. The results are shown in Figure~\ref{F:353_compare}. If temperature variation were the main cause of the residuals, we would expect to find a positive correlation between this map and our residual map. However, visual inspection already suggests there is no such correlation. 

Assuming that the dust emission follows the MBB spectrum, that all of the hydrogen gas shares the same MBB parameters as described in Equation~\eqref{E:MBB}, and that there is no spatial variations in the dust-to-gas ratio $\delta_{DG}$ or the spectral index $\beta$, the observed intensity at frequency $\nu$ can be expressed as:
\begin{equation}\label{E:MBB_1comp}
I_\nu^d(\nu,\hat{\theta}) = a N_{{\rm H}}(\hat{\theta})\nu^\beta B_\nu(\nu, T(\hat{\theta})) + b_\nu~~~,
\end{equation}
where $a$ is proportional to the dust-to-gas ratio $\delta_{\rm DG}$. If there exists a small spatial variation of temperature $T(\hat{\theta})$, we can expand the temperature around a fiducial mean temperature $\overline{T}$:
\begin{equation}
T(\hat{\theta}) = \overline{T} + \delta T(\hat{\theta})~~~.
\end{equation}
The slope of the relation between the two residual maps in Figure~\ref{F:353_compare} can then be calculated to the first order perturbation on $T(\hat{\theta})$:
\begin{equation}
\begin{split}
\frac{\delta I_\nu^{\nu_t}}{\delta I_\nu^H}
&=\frac{I_\nu^d(\hat{\theta})-I_\nu^{\nu_t}(\hat{\theta})}{I_\nu^d(\hat{\theta}) - I_\nu^H(\hat{\theta})}\\
&=1 - \left(\frac{\nu_t}{\nu}\right)\frac{1-e^{-h\nu/k_B\overline{T}}}{1-e^{-h\nu_t/k_B\overline{T}}}~~~,
\end{split}
\end{equation}
where $\nu$ and $\nu_t$ are 857 and 353\,GHz, respectively. See Appendix~\ref{A:T_var} for detailed derivations. Assuming a fiducial mean temperature $\overline{T}=20$ K, we derive the slope $\delta I_\nu^{\nu_t}/\delta I_\nu^H=0.37$, shown as the blue dashed line in Figure~\ref{F:353_compare}.

If temperature variation were the dominant source of residuals, the ratio of two residual maps shown in Figure~\ref{F:353_compare} would follow the blue dashed line. Instead, we find that, if anything, these two maps have a weak anti-correlation, disfavoring pure temperature variations. This weak negative correlation could be the result of noise fluctuations and other systematics that cannot be easily quantified, so we do not interpret it physically. Future investigation is warranted.

We perform a similar test using the 100\,$\mu$m map from IRAS, where a steeper slope is expected given the non-linear dependence of $I_\nu$ on $T$ closer to the peak of the emission. Specifically, we fit the 353\,GHz map to the 100\,$\mu$m map instead of the 857\,GHz map, and compare the residual of this fit to the residual of fitting our multi-phase H templates to the 857\,GHz map. Similar analytical calculations give the expected slope to be $3.48$ (see Appendix~\ref{A:T_var} for detailed derivations), almost an order of magnitude larger than the comparison between 353 and 857\,GHz in Figure~\ref{F:353_compare} ($0.37$). We use the Improved Reprocessing of the IRAS Survey (IRIS) 100\,$\mu$m map \citep{2005ApJS..157..302M}, which is a reprocessed version of the IRAS map at 100\,$\mu$m \citep{1984ApJ...278L...1N} that improves the zodiacal light subtraction, calibration, and destriping compared to the original IRAS map\footnote{\url{https://lambda.gsfc.nasa.gov/product/foreground/fg_iris_get.html}}.

Despite IRIS employing a zodiacal light removal process, fitting our multi-phase H templates to the IRIS map results in residuals exhibiting a strong gradient toward the ecliptic plane. Clearly some zodiacal light contamination persists and indeed dominates at moderate and low ecliptic latitudes. Therefore, we restrict our analysis to regions with ecliptic latitude $>50^\circ$, reducing the sky fraction to only $3.7\%/2.6\%$ in the NGC/SGC, respectively. We find a strong negative correlation between the two residual maps, inconsistent with expectations from temperature fluctuations. It is likely that, despite an aggressive ecliptic plane mask, zodiacal light still contributes to errors in our fitting. Therefore, we consider the results to be highly subject to systematics that cannot be well controlled.

Despite multiple lines of analysis, we find no evidence that dust temperature variations are driving residuals in the relation between $N_{\rm HI}$ and dust emission at high Galactic latitudes, with the exception of a single isolated region (Region~D). Indeed, on the basis of the tight correlation of the dust emission maps with $N_{\rm HI}$, we argue that dust temperature variations inferred from parametric fits to the FIR emission have been overstated, 

\subsection{Dust Opacity Law Variation}\label{S:beta_var}
Variations in the opacity law $\kappa_\nu$, parameterized as a power law $\kappa_\nu \propto \nu^\beta$ in the MBB model, can result in frequency-dependent residuals, similar to those from temperature variations. Unlike $T$ or $\delta_{\rm DG}$, $\beta$ is a parameter describing the relationship between two or more frequency maps: a positive or negative $\beta$ fluctuation makes no definite prediction for $I_\nu/N_{\rm H}$ at a single frequency. In addition to this complication, $\beta$ and $T$ have largely degenerate effects on the dust frequency spectrum \citep{Shetty:2009}. Previous studies from Planck have found weak evidence of $\beta$ variations. For instance, \cite{planck2014-a12} fitted an all-sky model of Galactic components, including thermal dust emission described by the MBB model, to the Planck data, and found a quite uniform distribution of $\beta$, with no detected spatial variation given their level of noise and systematics.



On the other hand, a recent analysis of SPIDER data found significant $\beta$ variations within the 4.8\% of sky analyzed in the Southern Galactic hemisphere \citep{2024arXiv240720982S}. One region within the SPIDER footprint exhibits an anomalously low $\beta$ of $1.09\pm0.09$, which is inconsistent at $3.9\sigma$ with the $\beta=1.52\pm0.06$ found over the rest of the SPIDER footprint. This higher value is consistent with the all-sky determination from \citet{planck2016-l11A} ($\beta=1.53\pm0.02$). The SPIDER low-$\beta$ region overlaps roughly with our Region~B, where we also observe large positive residuals. This suggests that variations in dust properties may still exist in certain regions of the sky.

The degeneracy between $\beta$ and $T$ complicates interpretation of multi-frequency analyses, while absolute degeneracy between the dust opacity at a fixed frequency and $\delta_{\rm DG}$ (see Equation~\eqref{eq:theory}) makes it impossible to attribute residuals to one or the other without ancillary data. For now, we must remain open to the possibility that the composition of dust may be variable in the Galactic ISM, affecting both FIR dust emission as well as optical extinction per unit dust mass and thus per $N_{\rm H}$.

\subsection{Magnetic Field Orientation}\label{S:polarization}
Dust grains emit polarized radiation because they are both aspherical and systematically aligned, with the grain short axis tending to be parallel to the local magnetic field. As a consequence of the alignment of aspherical grains, the effective extinction cross section of dust changes as the orientation of the magnetic field changes relative to the observer \citep{Lee:1985, 2019ApJ...887..159H}. Specifically, the effective cross section is larger when the magnetic field (and thus short axis) is oriented along the line of sight and smaller when the magnetic field (and thus short axis) is in the plane of the sky. 

The observed intensity $I_\nu^d(\hat{\theta})$ is related to the polarized intensity $P_\nu(\hat{\theta})$ by \citep{2019ApJ...887..159H}
\begin{equation}
I_\nu^d(\hat{\theta}) = \left[\sum_{i}^{N_c} \epsilon_{\nu, i} N_{{\rm H}, i}(\hat{\theta})\right] + b_\nu - P_\nu(\hat{\theta}),
\end{equation}
where the first term in brackets represents the total intensity from different gas components (including gas in all phases). The offset term $b_\nu$ accounts for contributions not related to Galactic dust, such as instrumental zero-level and the extragalactic background. A negative correlation between intensity and polarized intensity is expected, since when the magnetic field is along the line of sight, the polarized intensity goes to zero, while the asymmetric dust grains have a larger effective cross section, resulting in higher emissivity.

The polarization fraction $p_\nu(\hat{\theta})$ is defined as
\begin{equation}
p_\nu(\hat{\theta}) = \frac{P_\nu(\hat{\theta})}{\left[\sum_{i}^{N_c} \epsilon_{\nu, i} N_{{\rm H}, i}(\hat{\theta})\right]} = \langle p_\nu \rangle + \delta p_\nu(\hat{\theta}),
\end{equation}
where here we express it as the sum of a mean value $\langle p_\nu \rangle$ and a spatial fluctuation term $\delta p_\nu(\hat{\theta})$. Over a wide range of $N_{\rm H}$, $\langle p_\nu \rangle \simeq 5\%$ \citep{planck2014-XIX}. Therefore, the observed intensity can be written as
\begin{equation}
\begin{split}
I_\nu^d(\hat{\theta}) = &\left[\sum_{i}^{N_c} \epsilon_{\nu, i} N_{{\rm H}, i}(\hat{\theta})\right]\left(1 - \langle p_\nu \rangle\right) + b_\nu\\
&- \left[\sum_{i}^{N_c} \epsilon_{\nu, i} N_{{\rm H}, i}(\hat{\theta})\right]\delta p_\nu(\hat{\theta}).
\end{split}
\end{equation}

When we fit $\epsilon_{\nu,i}$ to the data without taking into account the magnetic field orientation effect, the polarization fraction biases the inferred $\epsilon_{\nu,i}$ by a factor of $1-\langle p_\nu \rangle$, assuming the polarization fraction is uncorrelated with the dust and gas column density maps. Consequently, the modeled intensity map becomes
\begin{equation}
I_\nu^m(\hat{\theta}) = \left[\sum_{i}^{N_c} \epsilon_{\nu, i} N_{{\rm H}, i}(\hat{\theta})\right]\left(1 - \langle p_\nu \rangle\right) + b_\nu.
\end{equation}
Thus, we derive the following expression:
\begin{equation}\label{E:ratio_polfrac}
\frac{I_\nu^d(\hat{\theta})-b_\nu}{I_\nu^m(\hat{\theta})-b_\nu}-1=-\frac{\delta p_\nu(\hat{\theta})}{1-\langle p_\nu \rangle},
\end{equation}
where the left-hand side represents the fractional residual of our template-based model. If spatially varying magnetic field orientation were the primary cause of the residuals, the fractional residual would be negatively proportional to $\delta p_\nu$, with a slope close to unity, given that $\langle p_\nu \rangle$ has a maximum value of $\sim$20\% \citep{planck2016-l11B}.

\begin{figure}[ht!]
\begin{center}
\includegraphics[width=\linewidth]{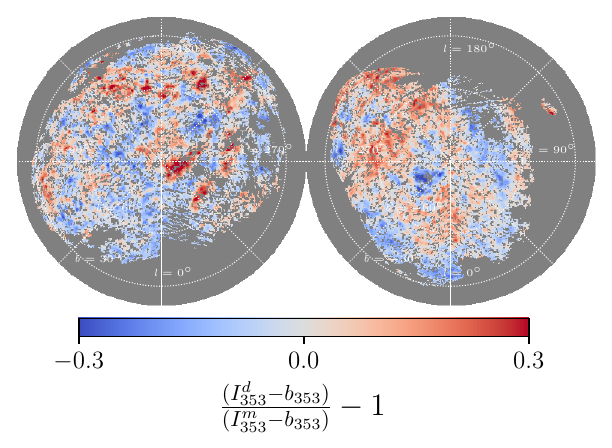}
\includegraphics[width=\linewidth]
{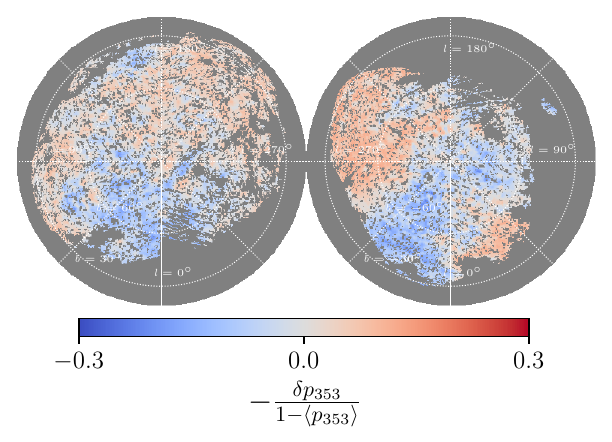}
\caption{\label{F:pol_map} Top: fractional residual of fitting our multi-phase H templates to the Planck 353\, GHz map. Bottom: negative of polarization fraction fluctuations divided by $1-\langle p_\nu\rangle$. The correlation coefficients in the two maps are $0.24$ (NGC) / $0.41$ (SGC). If magnetic field orientation effects were the only source of fitting residuals, the two maps would match with each other.}
\end{center}
\end{figure}

To test this, we construct the Planck polarization fraction map in the 353\,GHz band from the Planck~PR3 data, following the prescription outlined in \citet{planck2016-l11B}. Figure~\ref{F:pol_map} shows the 353\,GHz fractional residual and the polarization fluctuation (scaled by $1/(1-\langle p_\nu \rangle)$), i.e., the left- and right-hand sides of Equation~\eqref{E:ratio_polfrac}, respectively. Since the polarization maps from \citet{planck2016-l11B} are constructed at $80'$ resolution, we smooth our residual map to the same angular scale. The two maps have a correlation coefficient of 0.24 (NGC) / 0.41 (SGC). However, upon visual inspection, the large-scale patterns of the two maps are not well matched. The higher correlation in the SGC can be attributed to the correspondence in Region~B, where we observe strong positive residuals and lower fluctuations in polarization fraction. Nevertheless, neither the amplitude nor the detailed patterns of these two maps in Region~B match well. Additionally, we find a mean polarization fraction ($\langle p_\nu \rangle$) of $\sim10\%$, with $\sim5\%$ fluctuations ($\delta p_\nu$) in our high Galactic latitude regions. This is lower than the level needed to account for the residuals we observe at the $\sim20\%$ level. Therefore, we conclude that this is a subdominant effect. Given that \ion{H}{1} line intensity is insensitive to magnetic field orientation, this effect places a fundamental limit on the ability of gas-based tracers to predict dust emission.

\subsection{Extragalactic Contamination}\label{S:cmblensing}
In addition to emission from Galactic dust, the Planck maps contain dust emission from galaxies across a broad range of redshifts, the CIB. It is therefore possible that CIB fluctuations contribute to our model residuals. However, CIB fluctuations are expected to be at a much lower level than our large-scale residuals. According to a model of CIB fluctuations from \citet{2013A&A...557A..66B}, which has been validated with observations over the $100$--$1380$~$\mu$m range using multiple datasets, the CIB power spectra $C_\ell$ at $\ell=100$ are about $1 \times 10^4$/$1 \times 10^5$/$5 \times 10^5$\,Jy$^2$ sr$^{-2}$ at 353/545/857\,GHz, respectively. This corresponds to a degree-scale power of a few hundred Jy. However, our large-scale residual power exhibits fluctuations at a $\sim$10$^4$--$10^5$\,Jy level (see Figure~\ref{F:Planck_res}), much larger than can be explained by the CIB.

\begin{figure}[ht!]
\begin{center}
\includegraphics[width=\linewidth]{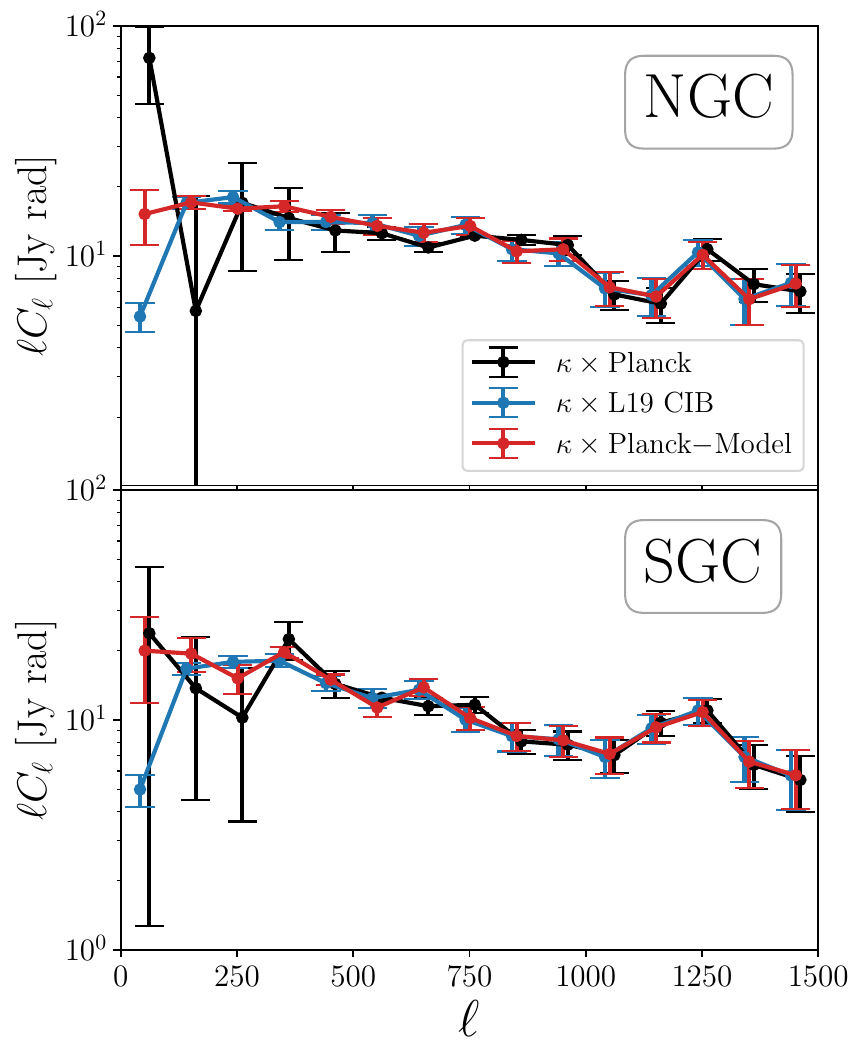}
\caption{\label{F:lens_res_xcorr} Cross power spectrum of the CMB lensing convergence and the Planck map (black), our residual map (red), and the CIB map from \citet{2019ApJ...883...75L} (blue) in the NGC (top) and the SGC (bottom).}
\end{center}
\end{figure}

The presence of CIB in the residuals can be probed through cross-correlation with other extragalactic tracers. Here, we investigate this correlation using the CMB lensing convergence map from Planck PR3 \citep{planck2016-l08}\footnote{\url{https://irsa.ipac.caltech.edu/data/Planck/release_2/all-sky-maps/maps/component-maps/lensing/COM_CompMap_Lensing_2048_R2.00.tar}}. The cross-power spectra between the lensing convergence and our residual map in 857\,GHz are displayed in Figure~\ref{F:lens_res_xcorr}. We also find similar results in 353 and 545\,GHz. Since the processed Planck map from \citet{2019ApJ...883...75L}, from which our residual map is derived, has been convolved with an additional beam function as part of the data processing to subtract the CMB (see Section~\ref{S:Planck_maps}), we multiply the resulting power spectrum by the beam window function provided by \citet{2019ApJ...883...75L}.

To account for the effects of masking, we employ the pseudo-$C_\ell$ estimator provided by the NaMaster package \citep{2019MNRAS.484.4127A}. For estimating the error on the binned power, we consider the Gaussian signal from sample variance and neglect mode coupling. Thus, the variance is given by 
\begin{equation}
\sigma_\ell^2=\frac{\left(C_\ell^{I\kappa}\right)^2+C_\ell^{II}C_\ell^{\kappa\kappa}}{\left(2\ell+1\right)f_{\rm sky}\Delta\ell},
\end{equation}
where $C_\ell^{I\kappa}$ represents the cross spectrum between the intensity and lensing convergence maps, $C_\ell^{II}$ and $C_\ell^{\kappa\kappa}$ correspond to the auto-spectra of the intensity and lensing convergence maps, respectively, $f_{\rm sky}$ denotes the fraction of sky coverage, and $\Delta\ell$ is the number of $\ell$ modes per bin \citep{1995PhRvD..52.4307K}.

For comparison, we also present the cross spectra with the $\kappa$ map using the Planck intensity map and the CIB map from \citet{2019ApJ...883...75L}, as shown in Figure~\ref{F:lens_res_xcorr}. The same beam window function correction is applied in these cross spectra. We observe good agreement in the cross-power spectra between the CMB lensing and the Planck map, our residual map, and the CIB map from \citet{2019ApJ...883...75L}. This confirms that the CIB is preserved in our residual maps.

We note that while \citet{2019ApJ...883...75L} fit the Galactic components with a linear model of 14 \ion{H}{1} velocity bins independently in each $N_{\rm side}=16$-sized super-pixel, which requires many more free parameters compared to our framework, and fluctuations at scales larger than their super-pixel size ($\ell\sim100$) are suppressed by construction, their approach does not result in absorbing some of the CIB signal into their Galactic components. This is evident in Figure~\ref{F:lens_res_xcorr} as the small-scale ($\ell\gtrsim100$) correlations with the $\kappa$ map are consistent with the correlation with our residuals and with the Planck map.

In summary, we detect CIB fluctuations in our residual map through cross-correlation with CMB lensing. The cross power spectrum amplitude on small scales at $\ell\gtrsim100$ is consistent with the cross-correlation with the Planck map, suggesting that the CIB fluctuations are preserved in our residual map rather than in our dust template. This is expected, as the gas-based template is not expected to contain CIB contamination. We also confirm that the CIB map from \citet{2019ApJ...883...75L} retains the CIB fluctuations on $\ell\gtrsim100$ scales, despite their dust model involving many more free parameters than ours. Although the CIB is present in our residual map, its fluctuations on degree scales are far too small to explain the large-scale fluctuations.


\section{A New Dust Extinction Map}\label{S:data_release_sfd}

\subsection{Map Construction and Validation}
\begin{figure}[ht!]
\begin{center}
\includegraphics[width=\linewidth]{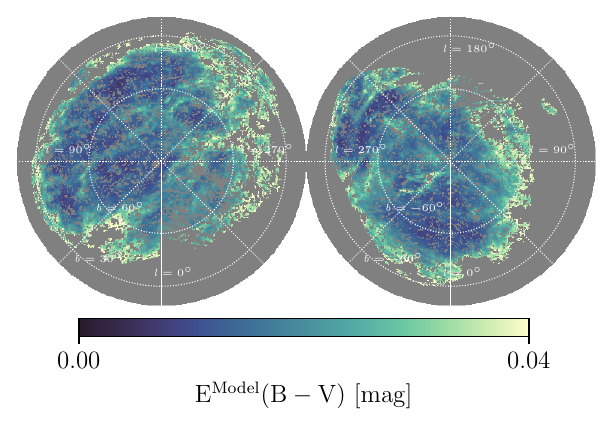}
\includegraphics[width=\linewidth]
{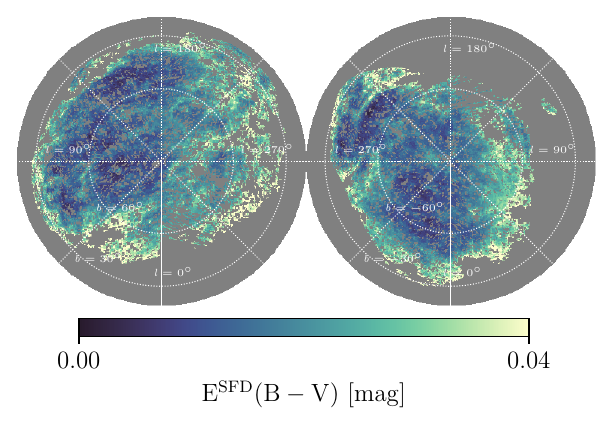}
\caption{\label{F:SFD_template} Top: the SFD reddening map. Bottom: our derived $E(B-V)$ model by fitting of fiducial template set to the SFD map. Both maps are smoothed with a $1^\circ$ Gaussian kernel to highlight the large-scale patterns.}
\end{center}
\end{figure}

\begin{figure}[ht!]
\begin{center}
\includegraphics[width=\linewidth]{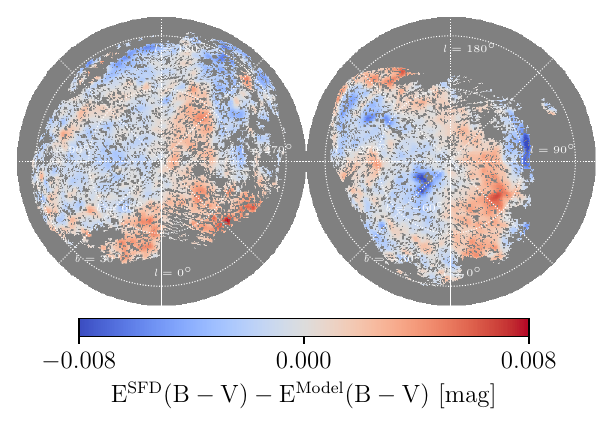}
\includegraphics[width=\linewidth]
{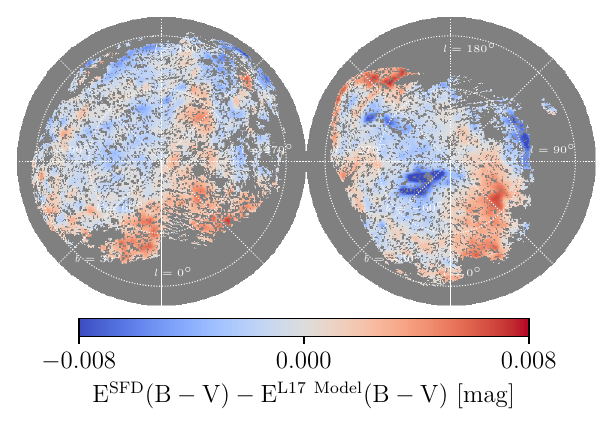}
\caption{\label{F:SFD_res_L17} Top: residual of fitting our multi-phase H templates to the SFD reddening map. Bottom: residual of fitting the single-component template from \citet{2017ApJ...846...38L} (i.e., \ion{H}{1} within $|v|<90$ km s$^{-1}$) to the SFD map. Both maps are smoothed with a $1^\circ$ degree Gaussian kernel to highlight the large-scale patterns.}
\end{center}
\end{figure}

\begin{figure}[ht!]
\begin{center}
\includegraphics[width=\linewidth]
{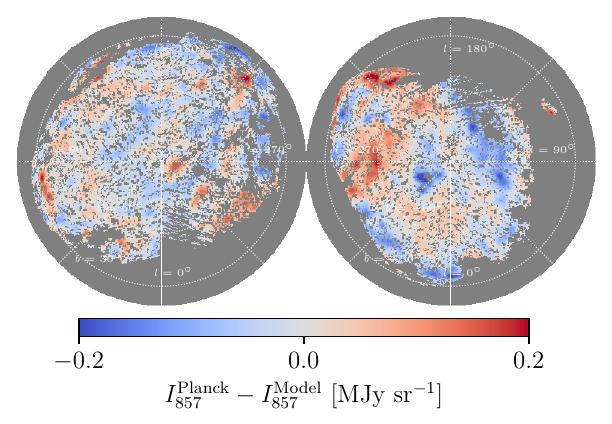}
\includegraphics[width=\linewidth]{figures/sfd_compare_sfd.pdf}
\includegraphics[width=\linewidth]{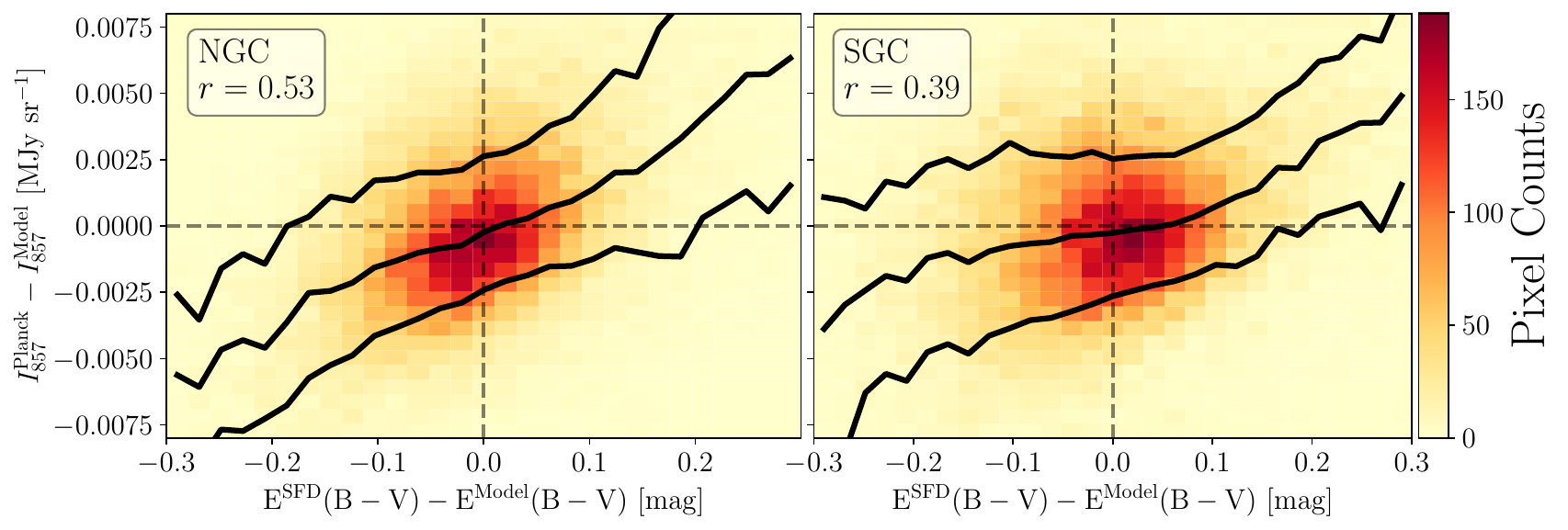}
\caption{\label{F:SFD_compare} Top: residual of fitting our multi-phase H templates to the Planck 857\,GHz map. Middle: residual of fitting our multi-phase H templates to the SFD reddening map. Bottom: the correlation between the residuals of fitting our templates to 857\,GHz map to the SFD reddening map in the NGC (left) and SGC (right). Black solid lines mark the 16th, 50th, and 84th percentiles. The Pearson correlation coefficient $r$ is noted in the box.}
\end{center}
\end{figure}

\begin{figure}[ht!]
\begin{center}
\includegraphics[width=\linewidth]{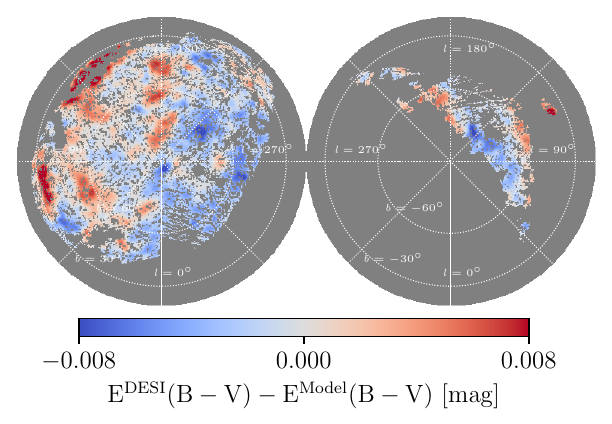}
\includegraphics[width=\linewidth]
{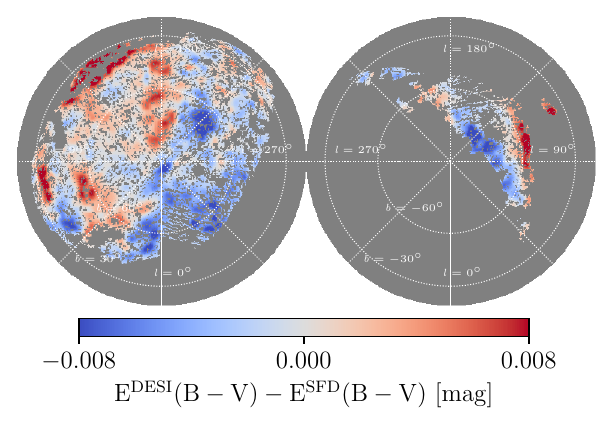}
\caption{\label{F:SFD_DESI} Top: residual of fitting our multi-phase H templates to the DESI stellar reddening map. Bottom: residual of fitting the SFD map to the DESI stellar reddening map.}
\end{center}
\end{figure}

With our multi-phase gas template set, we produce and release a new Galactic reddening map by fitting our templates to the SFD reddening map\footnote{\url{https://dataverse.harvard.edu/dataset.xhtml?persistentId=doi:10.7910/DVN/EWCNL5}} \citep{1998ApJ...500..525S}. Based on the FIR dust emission map at 100\,$\mu$m from IRAS, the construction of the SFD map included a zodiacal light subtraction and a temperature correction using 100 and 240\,$\mu$m maps from DIRBE, resulting in an all-sky reddening map with an angular resolution of $6'.1$. We employ the HEALPix representation of the SFD map with $N_{\rm side}=2048$. Subsequent analyses with photometry of stars from the Sloan Digital Sky Survey \citep{2010ApJ...725.1175S, 2011ApJ...737..103S} revealed a bias in the SFD calibration. Therefore, we have applied a multiplicative correction factor of $0.884$ from \citep{2011ApJ...737..103S} to the original SFD map. Figure~\ref{F:SFD_template} displays the SFD $E(B-V)$ map and our $E(B-V)$ model derived by fitting our templates to the SFD map. Specifically, we fit for the coefficients $\epsilon_{\rm SFD}$ and $b_{\rm SFD}$ in the following equation at the NGC and SGC, respectively:
\begin{equation}\label{E:linaer_model_SFD}
\begin{split}
E(B-V)_{\rm SFD}(\hat{\theta}) &= \left[\sum_{i}^{N_c} \epsilon_{{\rm SFD}, i} N_{{\rm HI}, i}(\hat{\theta})\right] \\
&+ \epsilon_{{\rm SFD},{\rm HII}} N_{{\rm HII}}(\hat{\theta}) + b_{\rm SFD}~~~.
\end{split}
\end{equation}

Figure~\ref{F:SFD_res_L17} displays the residual of fitting our template to the SFD map. For comparison, we also performed the fit with a previous \ion{H}{1}-based dust map from \citet{2017ApJ...846...38L}, which uses a single \ion{H}{1} template with a velocity threshold of $|v|<90$ km s$^{-1}$, shown in the bottom-left panel of Figure~\ref{F:SFD_res_L17}. Our model allows for different emissivities for different gas phases and \ion{H}{1} components, leading to improvement in the fit. However, the large-scale pattern of the residuals remains. \citet{2010ApJ...719..415P} and \citet{2022ApJS..260...17S} provided corrections to the SFD map based on direct measurements of extinction in galaxy or stellar spectra. Their correction maps exhibit a strong correlation with the residuals from fitting both our templates and the template from \citet{2017ApJ...846...38L} to the SFD map shown in Figure~\ref{F:SFD_res_L17}, consistent with the previous finding of \citet{2017ApJ...846...38L}.

The residuals of fitting our templates to the SFD map and the 857\,GHz map exhibit a positive correlation, as shown in the bottom-right panel of Figure~\ref{F:SFD_compare}. This is not surprising, as both maps are predominantly FIR thermal emission from dust, despite the SFD map undergoing additional processing such as temperature correction. However, some strong features in the Planck residual map are not shared by the SFD residual map, such as the positive feature in Region~F and the negative feature in Region~D.

Stellar reddening measurements provide a direct probe of the extinction. However, their sensitivity and resolution are limited by stellar density at high Galactic latitudes, and thus the FIR emission or \ion{H}{1} maps are still the most commonly used extinction templates for extragalactic studies. As the stellar reddening measurement from DESI \citep{2024arXiv240905140Z} directly maps the Galactic extinction, we perform a fit with our multi-phase gas templates as well as the SFD map to the DESI reddening map. The residuals are shown in Figure~\ref{F:SFD_DESI}. Our templates result in a residual fluctuation at the level of 3.5 (NGC)/3.0 (SGC) mmag, whereas the fit with SFD gives a higher residual of 4.2 (NGC)/3.1 (SGC) mmag. This suggests that our templates can serve as a more accurate dust template than the SFD map. The two residuals shown in Figure~\ref{F:SFD_DESI} share a very similar large-scale pattern, indicating that the remaining residuals are likely due to systematics in photometric calibration in the DESI reddening map. Finally, we note that while the DESI map is a direct extinction measurement, our dust map is built by fitting to the SFD map in order to access the full coverage of both the NGC and SGC fields. We find that fitting our templates to DESI and to SFD results in very similar best-fit coefficients in the NGC, where both maps have full coverage, thus validating that the large-scale systematics in the SFD map do not affect our final dust map.

\begin{figure}
\begin{center}
\includegraphics[width=\linewidth]{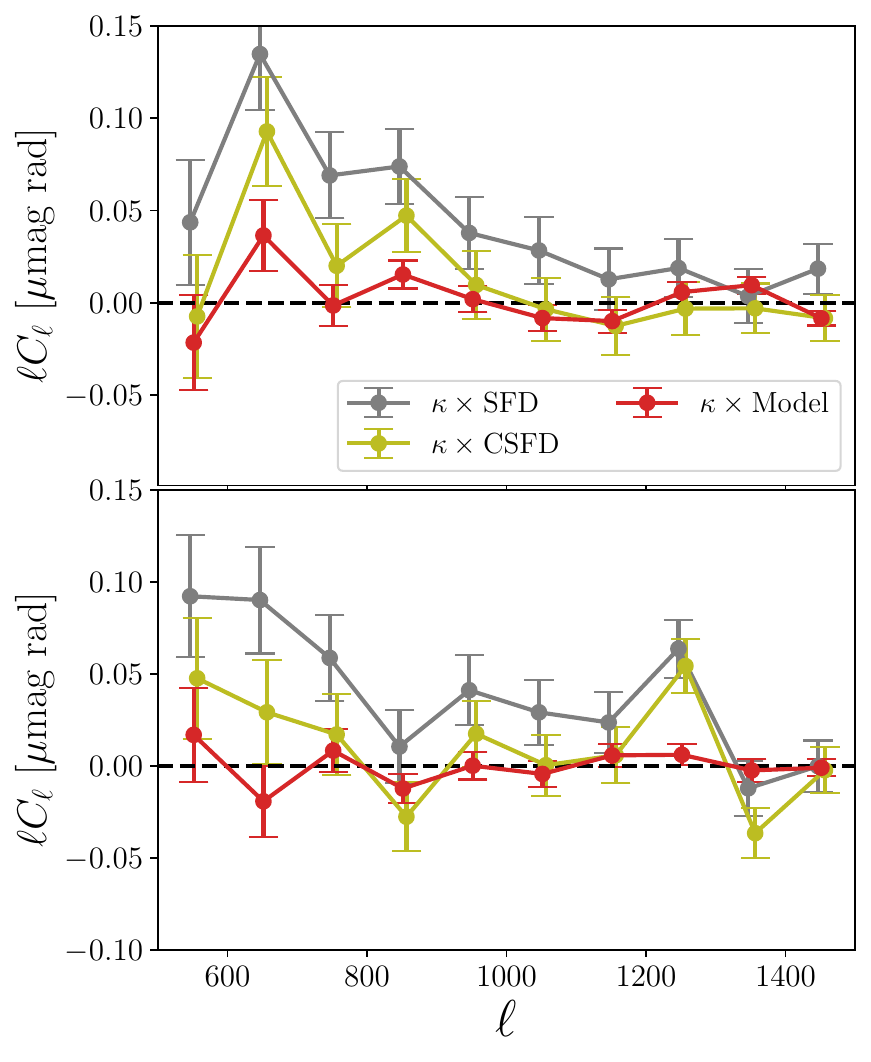}
\caption{\label{F:xcorr_model_sfd}Cross power spectrum of the CMB lensing convergence $\kappa$ and the reddening maps from SFD (gray), CSFD (yellow), and our model (red) in the NGC (left) and the SGC (right). We have conducted a high-pass filtering for $\ell<500$ modes on all three reddening maps to reduce the large-scale power leakage to the small-scale modes, and thus we only show the modes with $\ell>500$.}
\end{center}
\end{figure}

Recently, \citet{2023arXiv230603926C} constructed a corrected SFD map (CSFD) by deprojecting extragalactic contamination in the SFD map using cross-correlations with spectroscopic galaxy samples. As shown in Figure~\ref{F:xcorr_model_sfd}, on small scales ($\ell \gtrsim 500$), the SFD map contains CIB fluctuations that result in a positive correlation with the CMB lensing convergence ($\kappa$) map, whereas this correlation is significantly reduced in the CSFD map. We also calculate the cross spectrum of our $E(B-V)$ map and the $\kappa$ map, and, as expected since our model is based on Galactic gas tracers that are free from extragalactic signals, this correlation is consistent with zero. If we fit our templates to the CSFD map rather than to the SFD map, the resulting residuals display a similar large-scale fluctuation pattern as the SFD residual. This is not a surprising since the CIB exhibits negligible fluctuations on such large scales. Additionally, we note that the cross-correlation method employed in \citet{2023arXiv230603926C} predominantly constrains correlations at smaller angular scales and is thus not sensitive to possible large-scale extragalactic signals.

In summary, while our model improves upon previous \ion{H}{1}-based models from \citet{2017ApJ...846...38L} through the inclusion of both multiphase information and the clustering-based \ion{H}{1} template construction, the similar large-scale residual patterns with respect to FIR emission and to stellar extinction persist. Our attempt to exhaust information from the gas-based templates does not lead to significant improvements in modeling the dust emission, implying that the remaining residuals originate from factors other than those accounted for. Our investigation into the dust-to-gas ratio and temperature variations indicates that the residuals in some regions may be partly explained by these effects. However, there is no evidence that any single explanation can account for most of the residual pattern in our modeling.

\subsection{Data Product Release}
We make our new reddening map publicly available \footnote{\url{https://doi.org/10.7910/DVN/5E9EPW}}. This map has the same sky coverage as the analyses conducted throughout this study, covering 5577/4555\,deg$^2$ ($13.5\%/11.0\%$ of the sky) in the NGC/SGC. This map represents our best-fit model of linearly fitting our multi-phase template to the SFD map using Equation~\eqref{E:linaer_model_SFD} (with the $0.884$ multiplicative correction).

While the analysis presented in this work uses a reduced resolution of $N_{\rm side}=128$, we release the map at the original pixelization of the HI4PI data, which is $N_{\rm side}=1024$. We note that the \ion{H}{2} template is at a coarser pixelization of $N_{\rm side}=256$, and we upsample it to $N_{\rm side}=1024$. We release our reddening model as well as the template maps. The best-fit coefficients in Equation~\eqref{E:linaer_model_SFD} are $\{\epsilon_{\rm SFD, IVC-}, \epsilon_{\rm SFD, LVC}, \epsilon_{\rm SFD, HII}, b_{\rm SFD}\}$ $=\{8.74,10.51,1.51,1.22\}$ in the NGC and $\{\epsilon_{\rm SFD, IVC-}, \epsilon_{\rm SFD, LVC}, \epsilon_{\rm SFD, HII}, b_{\rm SFD}\}$ $=\{7.85,11.33,3.55,-0.70\}$ in the SGC with units mmag\,($10^{20}$\,cm$^{-2}$)$^{-1}$ and mmag for $\epsilon$'s and $b$'s, respectively.

In addition, we release the residual maps from fitting our model to the three Planck HFI frequency maps, as shown in the bottom row of Figure~\ref{F:Planck_res}. Like our reddening map product, these maps are upsampled to $N_{\rm side}=1024$ pixelization.


\section{Conclusions}\label{S:conclusions}
In this work, we build a new Galactic extinction map at high Galactic latitude covering 5577/4555\,deg$^2$ ($13.5\%/11.0\%$ of the sky fraction) in the NGC/SGC. Our extinction map is constructed with tracers of gas in neutral, ionized, and molecular phases. Utilizing \ion{H}{1} spectral maps, we adopt a data-driven approach to identify neutral gas in PPV space, segmenting it into distinct clouds to serve as fitting templates. This method is an improvement over previous \ion{H}{1}-based dust maps, which did not incorporate multi-phase information and relied solely on a simple velocity cut to define gas clouds. We assess the performance of our template by examining the residuals from fitting it to the Planck maps at 353, 545, and 857\,GHz. Additionally, we produce a reddening map by fitting our templates to the SFD reddening map.

Despite the enhanced analysis compared to previous \ion{H}{1}-based dust templates, we observe only a modest improvement in reducing the residuals when fitting to the Planck maps. We further investigate the origin of the remaining large-scale residuals using ancillary data that traces variations in different dust-to-gas properties in the ISM.

The principal conclusions of this work are as follows:
\begin{enumerate}
    \item At high-latitude regions, including the \ion{H}{2} template in addition to \ion{H}{1} reduces the residuals when fitting to Planck maps in some areas, but the improvement is modest.
    \item Currently available H$_2$ tracers built from CO emission lack the sensitivity needed to detect high-latitude molecular gas. Consequently, including the H$_2$ template does not help reduce the fitting residuals.
    \item Our new \ion{H}{1} templates, constructed using a clustering algorithm, improve upon previous templates that were based on a simple velocity cut. The improvements are particularly notable in reducingresiduals toward sight lines that contain the Magellanic Stream, IVCs, or have a significant fraction of \ion{H}{2}, which exhibit large positive or negative residuals when using previous \ion{H}{1}-based dust templates.
    \item We find that most of the large-scale patterns in the residuals observed with previous \ion{H}{1}-based templates cannot be effectively removed by these improved templates.
    \item We find evidence that the remaining residuals in some regions are associated with variations in the dust-to-gas ratio, dust emissivity, temperature, and magnetic field orientation. However, none of these factors appear to be the dominant effect responsible for the bulk of the residuals observed in our fit.
    \item Considering a constant dust temperature along each line of sight, we derive an upper bound on the dust temperature variation of $\sigma_T<1.28$~K in our high-latitude fields by assuming the residuals are entirely contributed by temperature variation and adopting a conservative mean temperature of $\overline{T}=25$~K. Our limit is lower than many previous dust temperature maps, suggesting that systematics may exist in their estimations.
    \item We detect the CIB in the residuals of fitting our template to the Planck map through cross-correlation with the CMB lensing map. This is expected, as our gas templates only trace the Galactic component. However, the CIB fluctuations are confined to much smaller scales, whereas the large-scale patterns in our residuals span over tens of degrees cannot be explained by CIB fluctuations.
\end{enumerate}

There are a few potential future directions for improving dust templates. First, it is essential to understand the systematics in all tracer maps, as highlighted by \citet{2024ApJ...961..204S}. Improved methods for identifying gas structures \citep[e.g.,][]{2024A&A...681A...1A} could also potentially enhance the results. Furthermore, the recent 3D dust extinction maps can help identify distinct gas clouds with additional line-of-sight information \citep{2019ApJ...887...93G,2022A&A...664A.174V,2024arXiv240702859S,2024A&A...685A..82E,2024arXiv240714594Z,2024ApJS..272...20A,2024MNRAS.532.3480D}. A joint analysis using multiple tracers to simultaneously constrain variations in dust properties could significantly enhance our existing analysis. Our current work with multi-phase data provides a foundation for this type of joint analysis.

Future observations with different tracers, such as diffuse interstellar band (DIB) absorption features \citep{2023ApJ...954..141S,2024arXiv240901777L}, diffuse Galactic light (DGL) from scattered stellar emission, and polycyclic aromatic hydrocarbon (PAH) emission, could offer further insights into the ISM. The upcoming SPHEREx mission \citep{2014arXiv1412.4872D}\footnote{\url{http://spherex.caltech.edu}} is promising for mapping the DGL and PAHs across the sky. These new datasets could enhance our understanding of the variations in dust and gas properties in the ISM and thus to higher-fidelity mapping of Galactic extinction.

\section*{Acknowledgments}
We would like to thank the anonymous referee for
providing valuable comments that improved the manuscript. We thank Yi-Kuan Chiang, Susan Clark, Hans Kristian Eriksen, Shamik Ghosh, Eirik
Gjerl{\o}w, Sebastian Hutschenreuter, Daniel Lenz, Gina Panopoulou, Raphael Skalidis, Ingunn Wehus, and Rongpu Zhou for assistance with various data products and helpful conversations. This work is support by NASA ROSES grant 18-2ADAP18-0192. T.-C.C. acknowledges support by NASA ROSES grants 18-2ADAP18-0192 and 21-ADAP21-0122. Part of this work was done at Jet Propulsion Laboratory, California Institute of Technology, under a contract with the National Aeronautics and Space Administration (80NM0018D0004).

\software{
astropy \citep{2013A&A...558A..33A,2018AJ....156..123A,2022ApJ...935..167A}, scikit-learn \citep{2011JMLR...12.2825P},
HEALPix \citep{2005ApJ...622..759G}, healpy \citep{2019JOSS....4.1298Z}, NaMaster \citep{2019MNRAS.484.4127A}, dustmaps \citep{2018JOSS....3..695G}
}

\appendix

\section{Clustering Algorithm Implementation}\label{A:clustering}
We construct the \ion{H}{1} templates by separating the voxels into a few clusters, based on the notion that \ion{H}{1} gas within the same structure, i.e., clustered in the PPV space, has a similar dust emissivity per H atom. In the k-means clustering algorithm, each data point has $N_{\rm attr}$ attributes, and the clustering is performed in the $N_{\rm attr}$-dimensional parameter space. In our implementation, we treat each ``voxel'' in the \ion{H}{1} dataset as a data point.

The k-means clustering seeks the best assignment of the dataset into $k$ clusters that minimizes the variance of distances between clustering members and their cluster centers. Here, the distance is defined by the Euclidean distance in the $N_{\rm attr}$-dimensional parameter space. For the attributes, we use the angular coordinates, velocity, and the \ion{H}{1} channel density, defined as the \ion{H}{1} column density within the channel map divided by the channel velocity width ($\Delta N_{\rm H\text{\sc I}}^{\rm ch}/\Delta v^{\rm ch}$). While the correct distance measure for the angular space is the geodesic on the sphere, and there exist clustering algorithms designed to operate on a sphere \citep[e.g., ][]{JMLR:v6:banerjee05a}, we instead encode the angular coordinate into a 3D Cartesian coordinate. We project the 2D angle onto a unit sphere in 3D and find the corresponding Cartesian coordinates ($x$, $y$, and $z$). This enables an easier implementation with the standard k-means clustering algorithm. Therefore, in total, we have five attributes in our clustering task for HI4PI data ($N_{\rm attr}=5$) — spatial coordinates $x$, $y$, and $z$, velocity $v$, and the \ion{H}{1} channel density $\Delta N_{\rm H\text{\sc I}}^{\rm ch}/\Delta v^{\rm ch}$. We normalize each attribute by subtracting its mean and dividing by its standard deviation. In the following, $\tilde{x}$, $\tilde{y}$, $\tilde{z}$, $\tilde{v}$, and $\tilde{\rho}$ denote the five normalized attributes, respectively.

The k-means clustering algorithm seeks for optimal clustering assignment based on Euclidean distance $d_{ij}$ between data point $i$ and cluster center $j$ of the input attribute:
\begin{equation}
\begin{split}
d_{ij}^2 =& (\tilde{x}_i-\tilde{x}_i)^2 + (\tilde{y}_i-\tilde{y}_i)^2 + (\tilde{z}_i-\tilde{z}_i)^2 \\
&+ (\tilde{v}_i-\tilde{v}_i)^2 +(\tilde{\rho}_i-\tilde{\rho}_i)^2.
\end{split}
\end{equation}

We can apply a weighting to each attribute. This allows us to tune the relative contribution to the distance measures between attributes. With the weighting on attributes, the distance metric is defined as
\begin{equation}
\begin{split}
d_{ij}^2 =& w^2_x(\tilde{x}_i-\tilde{x}_i)^2 + w^2_y(\tilde{y}_i-\tilde{y}_i)^2 + w^2_z(\tilde{z}_i-\tilde{z}_i)^2 \\
&+ w^2_v(\tilde{v}_i-\tilde{v}_i)^2 +w^2_\rho(\tilde{\rho}_i-\tilde{\rho}_i)^2.
\end{split}
\end{equation}

\begin{figure*}
\centering  
\subfigure{\includegraphics[width=0.45\linewidth]{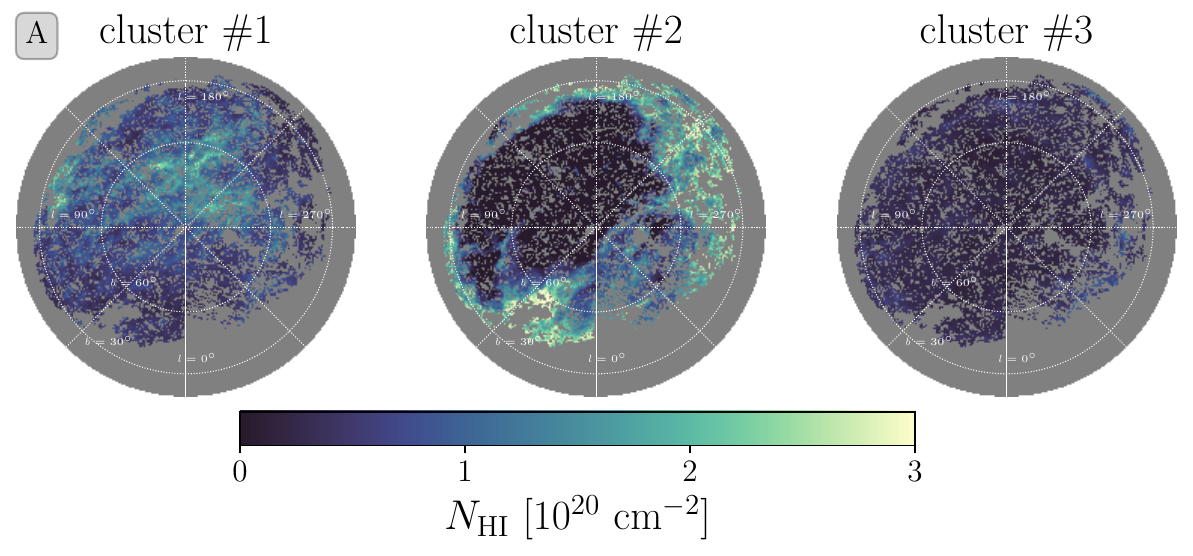}}
\subfigure{\includegraphics[width=0.45\linewidth]{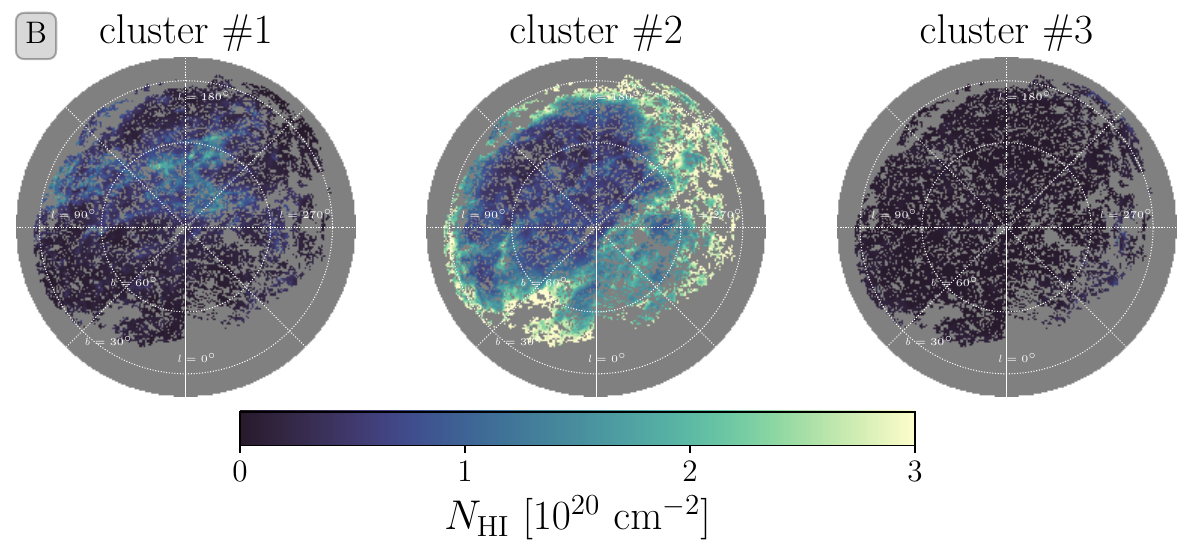}}
\subfigure{\includegraphics[width=0.48\linewidth]{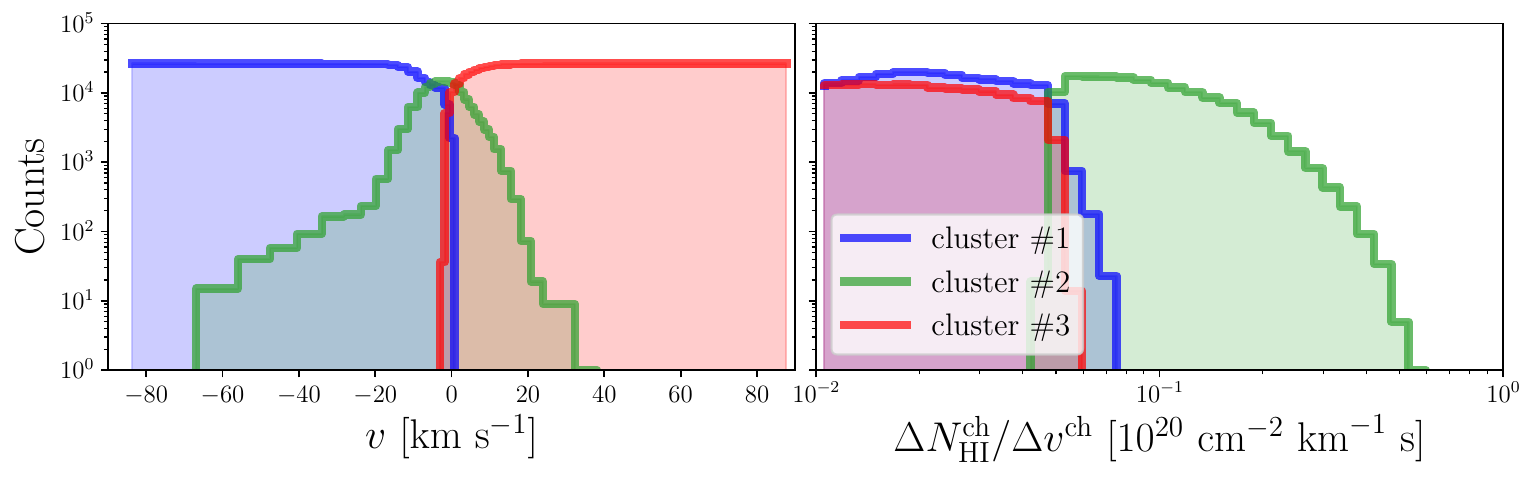}}
\subfigure{\includegraphics[width=0.48\linewidth]{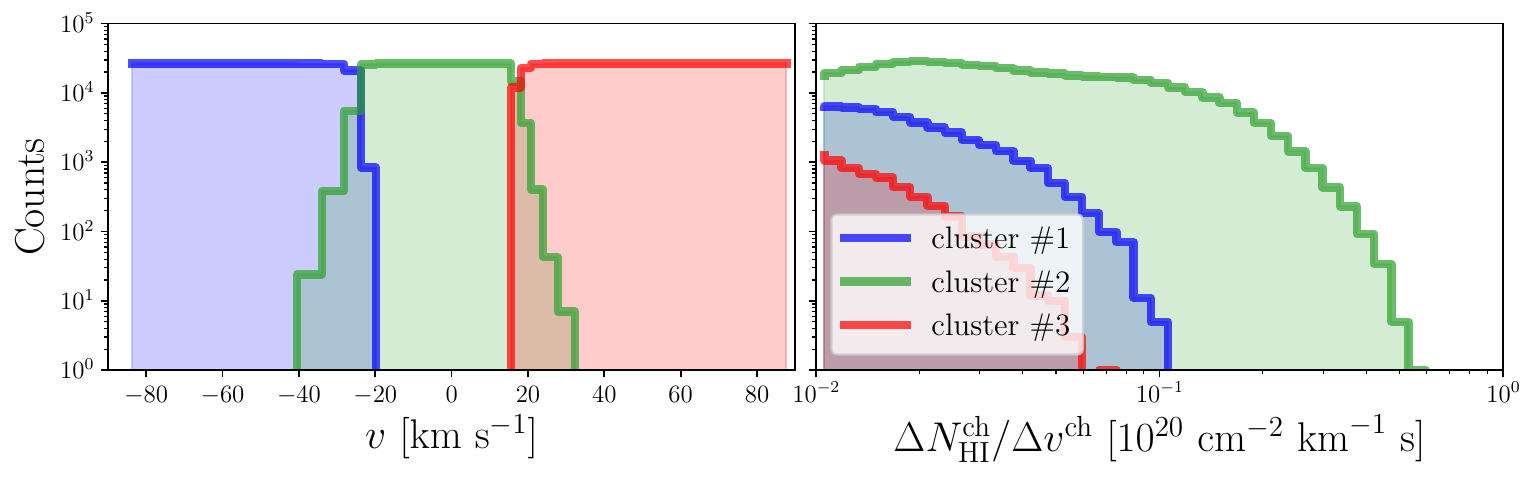}}
\subfigure{\includegraphics[width=0.45\linewidth]{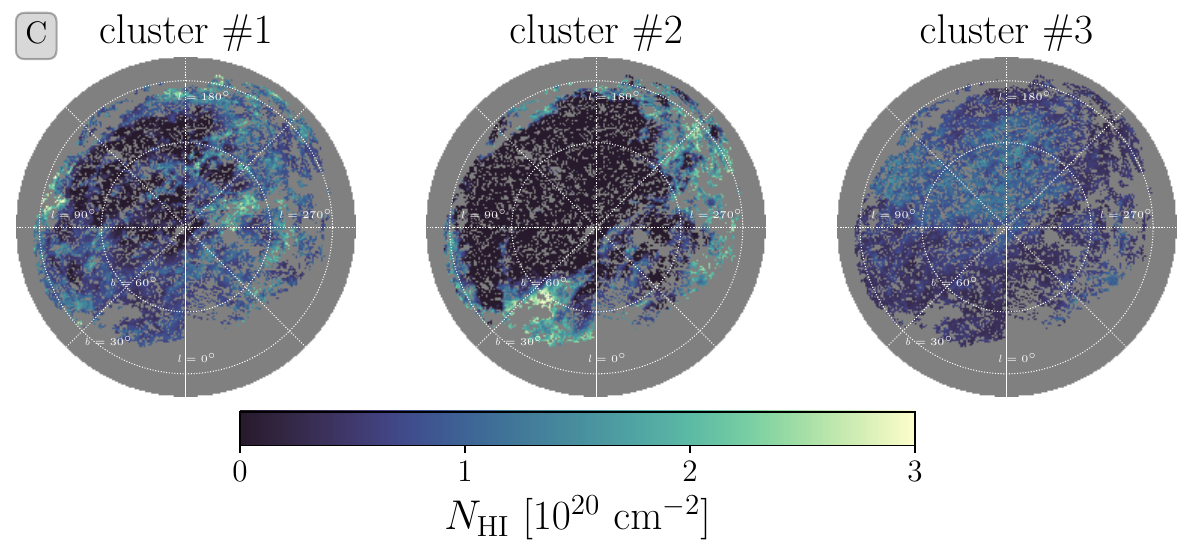}}
\subfigure{\includegraphics[width=0.45\linewidth]{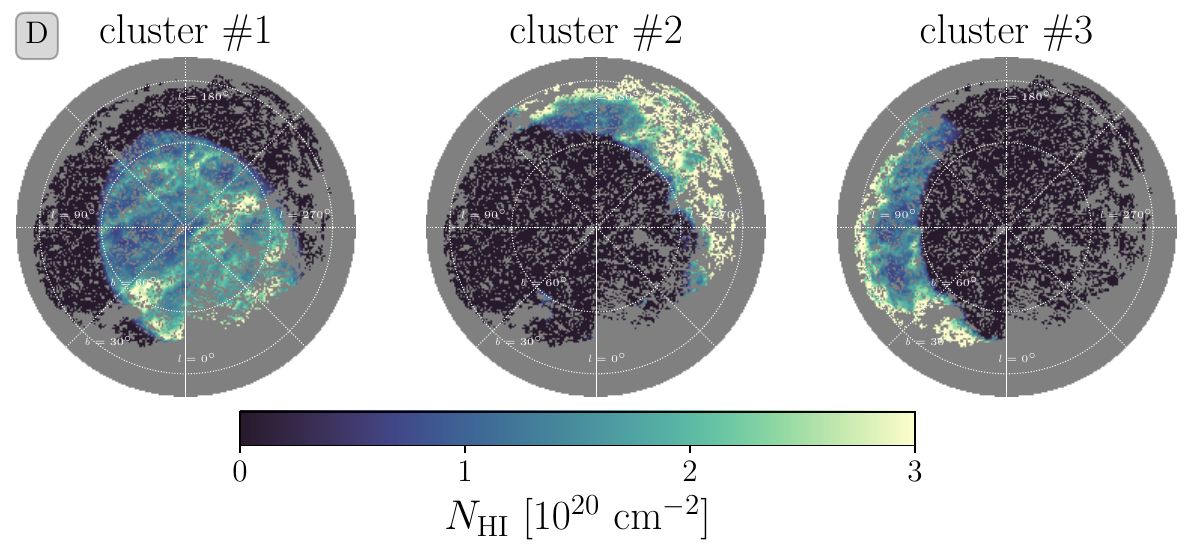}}
\subfigure{\includegraphics[width=0.48\linewidth]{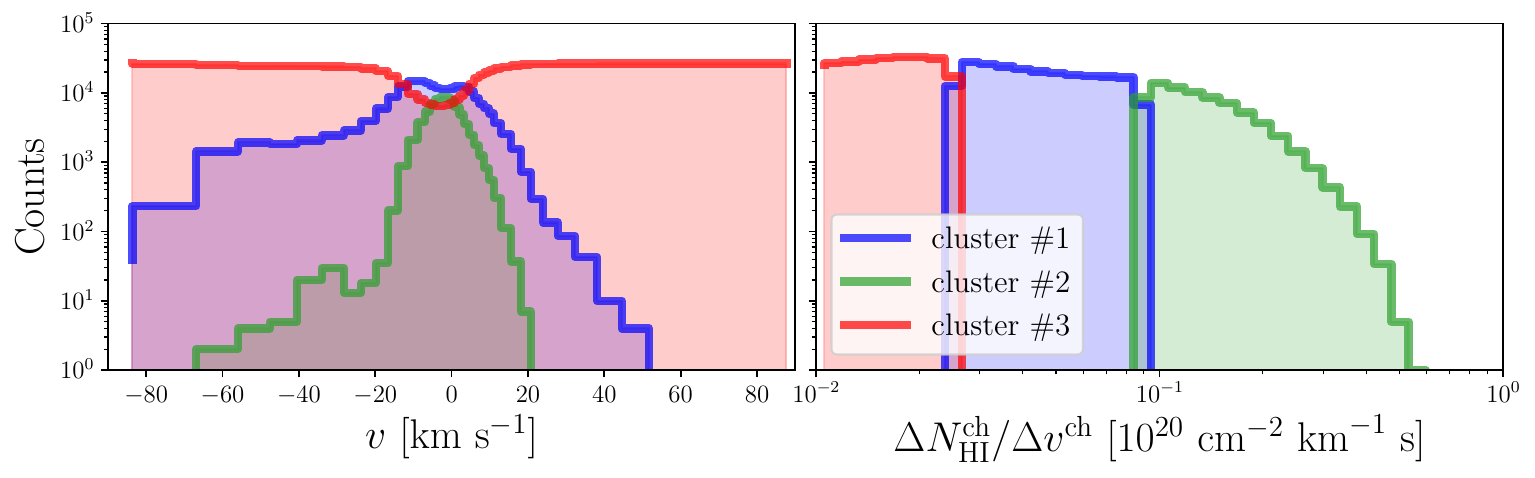}}
\subfigure{\includegraphics[width=0.48\linewidth]{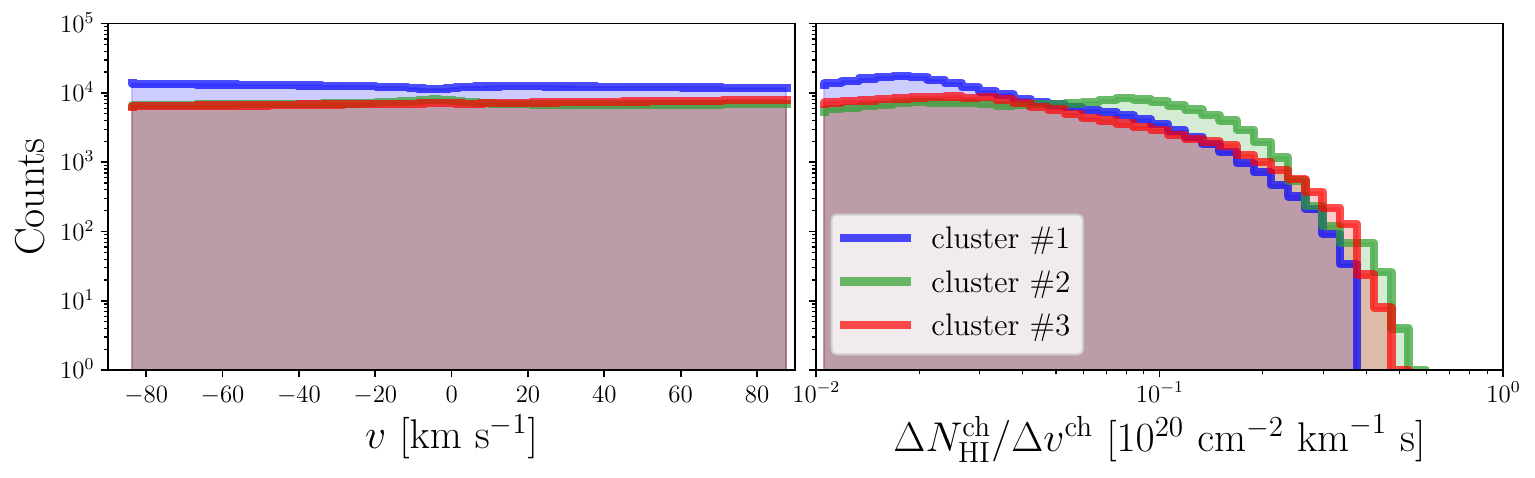}}
\caption{\label{F:clusdemo} Examples of results from applying the clustering algorithm on the NGC with three clusters using different attribute weightings. The maps display the column densities of the three resulting \ion{H}{1} template maps. The histogram panels are the velocity (left) and \ion{H}{1} channel density (right) histograms of the voxels for the three clusters. Panel~A: In this case, the velocity, \ion{H}{1} channel density, and spatial dimensions are balanced and thus, resulting clusters are partially separated in all three dimensions. Panel~B: In this case, we applied a large weighting on the velocity dimension, and thus the clusters show clear separation in velocity, but not the other dimensions. Panel~C: In this case, we applied a large weighting on the \ion{H}{1} channel density dimension, and thus the clusters show clear separation in \ion{H}{1} column density, but not the other dimensions. Panel~D: In this case, we applied a large weighting on the spatial dimension, and thus the clusters show clear spatial separation, but not for the other two dimensions.}
\end{figure*}

By symmetry, the weights are always set to the same value for $x$, $y$, and $z$. Since only the relative weights matter, there are two degrees of freedom for choosing the attribute weights: the velocity weight and the \ion{H}{1} channel density weight relative to the spatial coordinate weight. Therefore, without loss of generality, we set $w_x = w_y = w_z$ and explore the optimal values of $w_v$ and $w_\rho$ that provide the best fit to the data. As a demonstration, Figure~\ref{F:clusdemo} shows the clustering results in the NGC with three clusters ($k=3$) using different weightings. When applying a strong weight to one of the attributes, the algorithm tends to separate the data along that dimension as it has a much greater contribution to the distance measure. In our implementation, during both optimization for the number of clusters $k$ and our final model with the selected $k$, we search for the best attribute weighting ($w_v$ and $w_\rho$) that minimizes the $\chi^2$ when fitting to Planck data. For our final choice of number of clusters (3/5 clusters for NGC/SGC), the optimal weights are $w_v = 3.56$ and $w_\rho = 0.022$. This implies that the velocity of clouds is the most crucial information in classifying \ion{H}{1} into clusters, as expected, while the density of each 3D voxel is the least important feature.

We implement the k-means clustering algorithm with the \textit{cluster.KMeans} class in the \textit{scikit-learn} package \citep{2011JMLR...12.2825P}. We use the default Lloyd’s method of k-means implementation, and set the relative tolerance parameter $tol=0.1$. We use their ``greedy k-means$++$'' method for setting initial cluster centroids, and set the \textit{$n\_init$} parameter to 1, meaning that we only run the clustering once instead of multiple times with different initializing and pick the one with the best performance. We have tested that our results are consistent across multiple runs on different initialization.

\section{Fitting Residual from Temperature Fluctuations}\label{A:T_var}
Here, we provide detailed derivations of the ratio between different residual maps discussed in Section~\ref{S:T_var}. We consider only the effect of temperature fluctuations, ignoring other potential sources of residuals such as spatial variations of dust-to-gas ratio and zero-point calibration. Therefore, this calculation gives the expected ratio between different residual maps in the case that temperature fluctuations are the dominant source of the residuals. Furthermore, we make the following assumptions: first, the thermal emission from dust can be described by an MBB model with negligible variations in the spectral index $\beta$, and the temperature variation is small enough to be expressed as a first-order perturbation around a mean temperature $\overline{T}$, i.e., $T(\hat{\theta})=\overline{T} + \delta T(\hat{\theta})$. Second, we assume there is only one gas component $N_{\rm H}$ across the sky. This is a reasonable approximation, since we have found that most of the dust emission is associated with the LVC component in our regions. Specifically, $\sim75\%$ of the FIR emission is associated with the LVC (NGC) / the sum of LVC$+$ and LVC$-$ (SGC) component, when comparing $\epsilon\cdot\langle N_{\rm H}\rangle$ of each component in our fiducial model (Section~\ref{S:combined_model}), where $\epsilon$ is the best-fit emissivity given in Table~\ref{T:params}, and $\langle N_{\rm HI}\rangle$ in the mean column density of each template.

In Section~\ref{S:T_var}, we consider the residuals of fitting our multi-phase H template to the 857\,GHz map, and fitting the 353\,GHz map to the 857\,GHz and 100\,$\mu$m (3000\,GHz) maps. Here, we denote $\{\nu, \nu', \nu_t\} = \{857, 3000, 353\}$\,GHz, respectively. The intensity of the observed data in these three bands can be expressed as:
\begin{eqnarray}
I_\nu^d(\nu,\hat{\theta}) &= a N_{{\rm H}}(\hat{\theta})\nu^\beta B_\nu(\nu,T(\hat{\theta})) + b_\nu~~~,\\
I_\nu^d(\nu',\hat{\theta}) &= a N_{{\rm H}}(\hat{\theta})\nu'^\beta B_\nu(\nu', T(\hat{\theta})) + b_{\nu'}~~~,\\
I_\nu^d(\nu_t,\hat{\theta}) &= a N_{{\rm H}}(\hat{\theta}){\nu_t}^\beta B_\nu(\nu_t, T(\hat{\theta})) + b_{\nu_t}~~~.
\end{eqnarray}
When fitting the H template to the $\nu=857$\,GHz map, the best-fit model (ignoring uncertainties from noise fluctuations) is given by
\begin{equation}
I_\nu^H(\nu,\hat{\theta}) = a N_{{\rm H}}(\hat{\theta})\nu^\beta B_\nu(\nu, \overline{T}) + b_\nu~~~.
\end{equation}
Similarly, the best-fit models for fitting $\nu_t=353$\,GHz map to the $\nu=857$\,GHz and $\nu'=3000$\,GHz maps are
\begin{eqnarray}
I_\nu^{\nu_t}(\nu,\hat{\theta}) &= a N_{{\rm H}}(\hat{\theta})\nu_t^\beta B_\nu(\nu_t,T(\hat{\theta}))\frac{\nu^\beta B_\nu(\nu,\overline{T})}{\nu_t^\beta B_\nu(\nu_t,\overline{T})} + b_\nu~~~,\\
I_{\nu'}^{\nu_t}(\nu,\hat{\theta}) &= a N_{{\rm H}}(\hat{\theta})\nu_t^\beta B_\nu(\nu_t,T(\hat{\theta}))\frac{{\nu'}^\beta B_\nu(\nu',\overline{T})}{\nu_t^\beta B_\nu(\nu_t,\overline{T})} + b_{\nu'}~~~.
\end{eqnarray}

With a small temperature fluctuation $\delta T(\hat{\theta})$, we expand the Planck function to the first order:
\begin{equation}
\begin{split}
B_\nu(\nu, T(\hat{\theta})) &= B_\nu(\nu, \overline{T} + \delta T(\hat{\theta}))\\
&=B_\nu(\nu, \overline{T}) + \left.\frac{\partial B_\nu}{\partial T}\right|_{\nu,\overline{T}} \delta T(\hat{\theta})~~~,
\end{split}
\end{equation}
where
\begin{equation}
\left.\frac{\partial B_\nu}{\partial T}\right|_{{\nu},\overline{T}}=\frac{h\nu}{k_B\overline{T}}\frac{1}{1-e^{-h\nu/k_B\overline{T}}}B_\nu(\nu,\overline{T})~~~.
\end{equation}

The residuals of fitting the H template to the $\nu=857$\,GHz map and fitting the $\nu_t=353$\,GHz map to the $\nu=857$\,GHz and $\nu'=3000$\,GHz maps can thus be expressed as 
\begin{equation}
\begin{split}
\delta I_\nu^H &= I_\nu^d(\hat{\theta}) - I_\nu^H(\hat{\theta})\\
&= a N_{{\rm H}}(\hat{\theta})\nu^\beta \left.\frac{\partial B_\nu}{\partial T}\right|_{\nu,\overline{T}} \delta T(\hat{\theta})~~~,
\end{split}
\end{equation}

\begin{equation}
\begin{split}
\delta I_\nu^{\nu_t} &= I_\nu^d(\hat{\theta}) - I_\nu^{\nu_t}(\hat{\theta})\\
&= a N_{{\rm H}}(\hat{\theta})\nu^\beta \delta T(\hat{\theta})\\ 
&\cdot\left[
\left.\frac{\partial B_\nu}{\partial T}\right|_{\nu,\overline{T}} - \left.\frac{\partial B_\nu}{\partial T}\right|_{{\nu_t},\overline{T}}\frac{B_\nu(\nu,\overline{T})}{B_\nu(\nu_t,\overline{T})}
\right]~~~,
\end{split}
\end{equation}
and
\begin{equation}
\begin{split}
\delta I_{\nu'}^{\nu_t} &= I_{\nu'}^d(\hat{\theta}) - I_{\nu'}^{\nu_t}(\hat{\theta})\\
&= a N_{{\rm H}}(\hat{\theta}){\nu'}^\beta \delta T(\hat{\theta})\\ 
&\cdot\left[
\left.\frac{\partial B_\nu}{\partial T}\right|_{{\nu'},\overline{T}} - \left.\frac{\partial B_\nu}{\partial T}\right|_{{\nu_t},\overline{T}}\frac{B_\nu({\nu'},\overline{T})}{B_\nu(\nu_t,\overline{T})}
\right] ~~~.
\end{split}
\end{equation}

In Figure~\ref{F:353_compare}, we show the correlation between $\delta I_\nu^{\nu_t}$ and $\delta I_\nu^H$, and the expected slope of the correlation is given by
\begin{equation}
\begin{split}
\frac{\delta I_\nu^{\nu_t}}{\delta I_\nu^H}
&=\frac{I_\nu^d(\hat{\theta})-I_\nu^{\nu_t}(\hat{\theta})}{I_\nu^d(\hat{\theta}) - I_\nu^H(\hat{\theta})}\\
&=1 - \frac{B_\nu(\nu,\overline{T})}{B_\nu(\nu_t,\overline{T})}\frac{\left.\frac{\partial B_\nu}{\partial T}\right|_{{\nu_t},\overline{T}}}{\left.\frac{\partial B_\nu}{\partial T}\right|_{{\nu},\overline{T}}}\\
&=1 - \left(\frac{\nu_t}{\nu}\right)\frac{1-e^{-h\nu/k_B\overline{T}}}{1-e^{-h\nu_t/k_B\overline{T}}}~~~.
\end{split}
\end{equation}
With our assumed fiducial mean temperature $\overline{T}=20$\,K, we obtain $\delta I_\nu^{\nu_t}/\delta I_\nu^H=0.37$.

We performed a similar test with IRAS data by comparing $\delta I_{\nu'}^{\nu_t}$ and $\delta I_\nu^H$, in which case the expected slope of the correlation is given by
\begin{equation}
\begin{split}
\frac{\delta I_{\nu'}^{\nu_t}}{\delta I_\nu^H}
&=\frac{I_{\nu'}^d(\hat{\theta})-I_{\nu'}^{\nu_t}(\hat{\theta})}{I_\nu^d(\hat{\theta}) - I_\nu^H(\hat{\theta})}\\
&=\left(\frac{\nu'}{\nu}\right)^{\beta}\frac{B_\nu(\nu',\overline{T})}{B_\nu(\nu,\overline{T})}\left[\frac{B_\nu(\nu,\overline{T})}{B_\nu(\nu',\overline{T})}\frac{\left.\frac{\partial B_\nu}{\partial T}\right|_{{\nu'},\overline{T}}}{\left.\frac{\partial B_\nu}{\partial T}\right|_{{\nu},\overline{T}}}\right. \\
&- \left.\frac{B_\nu(\nu,\overline{T})}{B_\nu(\nu_t,\overline{T})}\frac{\left.\frac{\partial B_\nu}{\partial T}\right|_{{\nu_t},\overline{T}}}{\left.\frac{\partial B_\nu}{\partial T}\right|_{{\nu},\overline{T}}}
\right]\\
&=\left(\frac{\nu'}{\nu}\right)^{3+\beta}\frac{e^{h\nu/k_B\overline{T}}-1}{e^{h\nu'/k_B\overline{T}}-1}\\
&\cdot\left[\left(\frac{\nu'}{\nu}\right)\frac{1-e^{-h\nu/k_B\overline{T}}}{1-e^{-h\nu'/k_B\overline{T}}}\right.\\
&-\left.\left(\frac{\nu_t}{\nu}\right)\frac{1-e^{-h\nu/k_B\overline{T}}}{1-e^{-h\nu_t/k_B\overline{T}}}\right]~~~.
\end{split}
\end{equation}
Similarly, assuming $\overline{T}=20$\,K and $\beta=1.5$, we obtain $\delta I_{\nu'}^{\nu_t}/\delta I_\nu^H=3.48$.

\vspace{5pt}
\bibliography{HIdust, Planck_bib}{}
\bibliographystyle{aasjournal}

\end{document}